\documentclass[aps,reprint,amsmath,amssymb,superscriptaddress,prl,showkeys]{revtex4-2}
\usepackage{xcolor}
\usepackage{graphicx}
\usepackage{dcolumn}
\usepackage{bm}
\usepackage{natbib}

\usepackage{textcomp} 

\begin{document}

\preprint{APS/123-QED}

\title{Robust Interlayer Exciton Interplay in Twisted van der Waals Heterotrilayer on a Broadband Bragg Reflector up to Room Temperature}

\author{Bhabani Sankar Sahoo}
\affiliation{Institut für Physik und Astronomie, Technische Universit\"at Berlin, Hardenbergstrasse 36, 10623 Berlin, Germany}

\author{Shachi Machchhar}%
\affiliation{Institut für Physik und Astronomie, Technische Universit\"at Berlin, Hardenbergstrasse 36, 10623 Berlin, Germany}

\author{Avijit Barua}%
\affiliation{Institut für Physik und Astronomie, Technische Universit\"at Berlin, Hardenbergstrasse 36, 10623 Berlin, Germany}

\author{Martin Podhorský}%
\affiliation{Institut für Physik und Astronomie, Technische Universit\"at Berlin, Hardenbergstrasse 36, 10623 Berlin, Germany}

\author{Seth Ariel Tongay}%
\affiliation{Materials Science and Engineering, School for Engineering of Matter, Transport and Energy, Arizona State University, Tempe, USA}

\author{Takashi Taniguchi}%
\affiliation{Research Center for Materials Nanoarchitectonics, National Institute for Materials Science,  1-1 Namiki, Tsukuba 305-0044, Japan}

\author{Kenji Watanabe}%
\affiliation{Research Center for Electronic and Optical Materials, National Institute for Materials Science, 1-1 Namiki, Tsukuba 305-0044, Japan}

\author{Chirag Chandrakant Palekar}%
\email{c.palekar@tu-berlin.de}
\affiliation{Institut für Physik und Astronomie, Technische Universit\"at Berlin, Hardenbergstrasse 36, 10623 Berlin, Germany}

\author{Stephan Reitzenstein}%

\affiliation{Institut für Physik und Astronomie, Technische Universit\"at Berlin, Hardenbergstrasse 36, 10623 Berlin, Germany}

\date{\today}

\begin{abstract} \textcolor{red}{\textbf{}}
We report robust room temperature interlayer excitons in transition metal dichalcogenide heterostructures engineered via precise stacking orientation and twist-angle control. We integrate 2H-stacked MoSe${_2}$/${^1}$WSe${_2}$/${^2}$WSe${_2}$ heterotrilayer onto a chirped distributed Bragg reflector that acts as a backside mirror. This way, we fabricate a platform that hosts distinct heterotrilayer, heterobilayer, and homobilayer regions with enhanced excitonic features at elevated temperatures. Although the heterobilayer supports temperature-tunable singlet and triplet interlayer excitons, it exhibits low emission yield at 4 K. In comparison, the heterotrilayer shows remarkable excitonic properties, including pronounced band modulation, intervalley interlayer exciton transitions, and a tenfold photoluminescence enhancement along with a sevenfold increase in exciton decay time at cryogenic temperatures compared to the heterobilayer system. Temperature-dependent studies reveal intriguing interlayer exciton dynamics in the heterotrilayer, including the emergence of valley-polarized interlayer excitons, and the ability to maintain optical stability up to room temperature. Our results establish a clear strategy for engineering excitonic states across multilayer van der Waals heterostructures from 4 K to room temperature, providing a versatile platform for excitonic optoelectronics, quantum photonics, and tunable long-lived interlayer exciton states in scalable TMD heterostructures.

\end{abstract}


\maketitle


\begin{figure*}
	\centering\includegraphics[width=18cm]{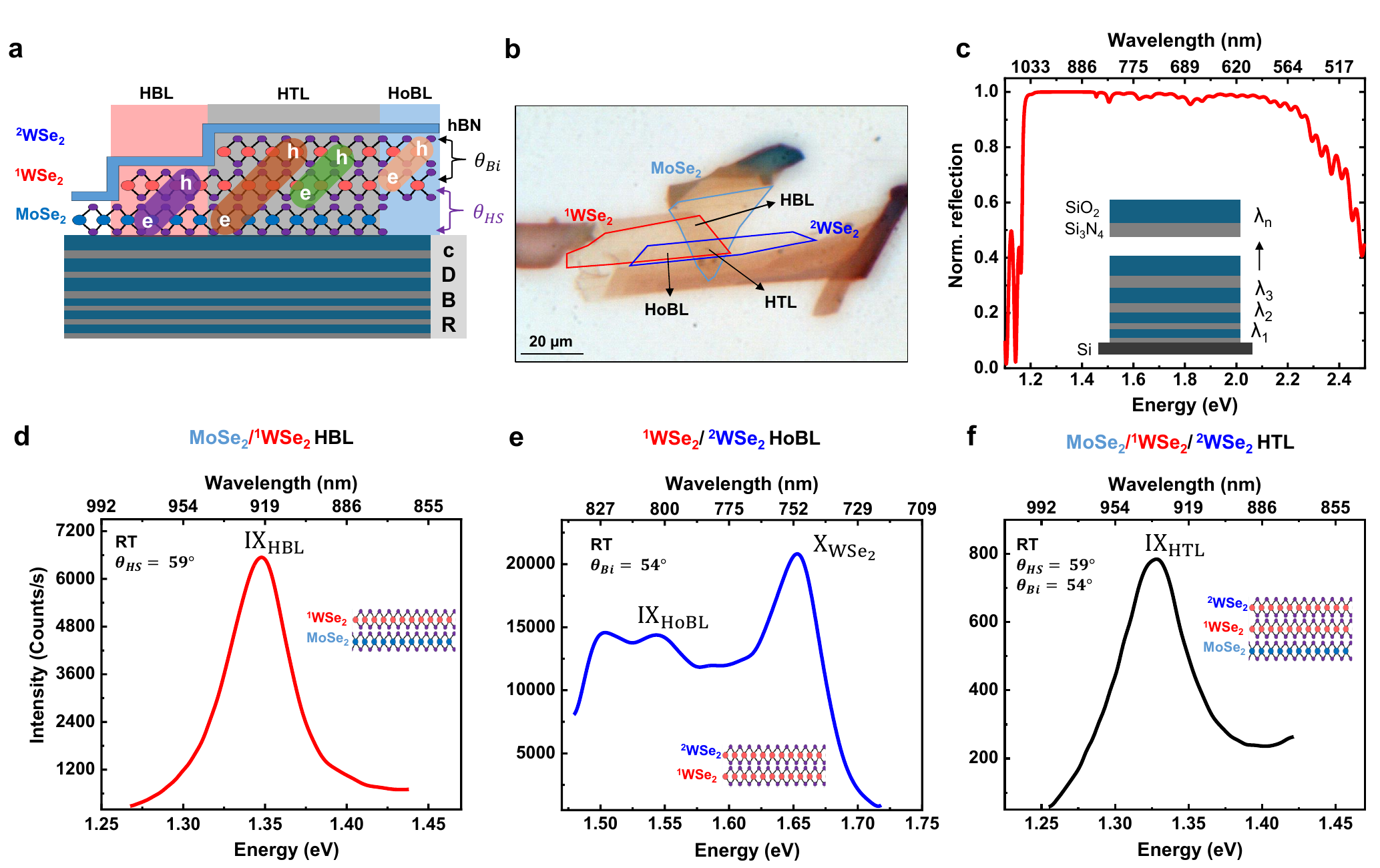} 
	\caption{\text{Structural and optical overview of twisted VdWHSs at RT.} \textbf{a} Schematic of the HTL layer design showing the stacking order of TMD monolayers and the substrate configuration along with IX formation between layers. \textbf{b} Optical microscope image highlighting the monolayers of MoSe\textsubscript{2}, \textsuperscript{1}WSe\textsubscript{2}, and \textsuperscript{2}WSe\textsubscript{2} with light blue, red, and blue outline respectively, indicating also the HTL, HBL, and twisted HoBL regions in the image. \textbf{c} Calculated reflection spectrum of the cDBR showing a 600 nm wide stopband. The inset illustrates the layer design of the cDBR. (\textbf{d}-\textbf{f}) RT PL spectra of the \textbf{d} MoSe\textsubscript{2}/\textsuperscript{1}WSe\textsubscript{2} HBL region \textbf{e} Twisted \textsuperscript{1}WSe\textsubscript{2}/\textsuperscript{2}WSe\textsubscript{2} HoBL region, and \textbf{f} MoSe\textsubscript{2}/\textsuperscript{1}WSe\textsubscript{2}/\textsuperscript{2}WSe\textsubscript{2} HTL region which excited by a continuous-wave (CW) laser with energy 1.87 eV (660 nm), revealing the distinct emission features associated with each stacking configuration.}
	\label{Figure_1}
\end{figure*}

\section{Introduction} \label{intro}

Two-dimensional (2D) transition metal dichalcogenides (TMDs) have attracted significant attention because of their strong Coulomb interactions, direct electronic band gaps in the monolayer limit, and pronounced excitonic effects \cite{gupta2025two, Chen2023, wangreview, Mak2010, Splendiani2010}. When different TMD monolayers are vertically stacked, they form van der Waals heterostructures (VdWHSs) that can host interlayer excitons (IXs), where electron-hole pairs are spatially separated across adjacent monolayers. These IXs exhibit long lifetimes, tunable emission energies, and valley selective optical properties, making them promising candidates for next generation optoelectronic devices \cite{Rivera2015, Jin2019, Alexeev2019, Seyler2019Nature}. These VdWHSs are particularly compelling because the optical and electronic properties of their IXs are highly sensitive to both the interlayer stacking angle and the stacking order. Variation in the twist angle modifies the interlayer coupling, momentum alignment, and hybridization of electronic states, leading to distinct excitonic resonances and the formation of moiré potential landscapes \cite{Palekar2024, Zhang2020, Tran2019}. In the special case of a homobilayer (HoBL), excitons confined within the same material generate intervalley and intravalley interlayer or intralayer transitions that contribute to photoluminescence (PL) emission, leading to complex spectral features  \cite{lindlau2018role, Brem2020}. When an additional monolayer is stacked on a heterobilayer (HBL), a heterotrilayer (HTL) structure is formed. This configuration provides additional degrees of freedom and combines the features of both HoBL and HBL. It hosts multiple excitonic transitions, such as interlayer, intralayer, and hybrid excitons, whose coupling and thermal stability has remained largely unexplored. Although recent studies have reported on basic features of HTLs \cite{Palekar2024_trilayer, Lian2023, Chen2022, Bai2022, Forg2021NatComm, Slobodkin2020}, a comprehensive understanding of how their optical properties differ from those of HBL, especially as a function of temperature, is still lacking, despite their high potential as active medium in advanced nanophotonic devices. In this context, temperature-dependent PL studies provide key insights into exciton decay times and valley polarization, offering a fundamental understanding of the interplay between radiative and non-radiative processes that governs exciton stability. Additionally, observing distinct PL signals at room temperature (RT) is particularly significant, as it demonstrates the robustness of IXs and the potential of VdWHS for practical optoelectronics and valleytronics applications. However, maintaining PL emission in VdWHS at elevated temperatures remains challenging due to enhanced non-radiative recombination.

In this work, we report on the fabrication and optical study of a MoSe\textsubscript{2}/WSe\textsubscript{2}/WSe\textsubscript{2} HTL on a multiresonant chirped distributed Bragg reflector (cDBR). This stacked TMD system naturally forms three distinct regions of HBL, HoBL, and HTL within a single VdWHS which facilitates comparative studies of its emission properties. The DBR's stopband was designed to spectrally overlap with the PL emission of the HBL, HoBL, and HTL regions, enabling efficient PL detection across visible to near infrared regions within a single resonator system \cite{Palekar2024_chirp}. In comprehensive optical studies on the realized VdWHS system, we performed temperature-dependent PL measurements and analyzed the evolution of radiative decay times as well as the valley polarization. A comparative analysis reveals clear differences in the temperature-dependent emission dynamics of the HBL and HTL, reflecting distinct exciton coupling mechanisms and a markedly enhanced thermal stability in the HTL system. Furthermore, temperature-dependent optical studies of the HBL, HoBL, and HTL regions uncover distinct excitonic and valley dynamics.

Overall, the observation of strong IX PL at RT across all three VdWHS regions demonstrates robust interlayer coupling and thermal stability enabled by the controlled twist angle and the backside cDBR. Notably, our results highlight the particular promise of the trilayer system for RT optoelectronic applications.


\section{Results} \label{Resutls}

We prepared a VdWHS sample consisting of stacked TMD monolayers by mechanical exfoliation \cite{Novoselov2004} and the dry transfer method \cite{Castellanos-Gomez2014}. During fabrication, each monolayer was stacked individually on the cDBR with a controlled twist angle between them. As illustrated in Figure \ref{Figure_1}(a), first a monolayer of MoSe\textsubscript{2} was stacked on the cDBR. Then a  monolayer of WSe\textsubscript{2} (\textsuperscript{1}WSe\textsubscript{2}) was vertically transferred onto the first layer with the aim of achieving a twist angle close to 0$^{\circ}$ or 60$^{\circ}$, forming a MoSe\textsubscript{2}/WSe\textsubscript{2} HBL. Finally, an additional WSe\textsubscript{2} (\textsuperscript{2}WSe\textsubscript{2}) monolayer was stacked on the HBL, forming a twisted WSe\textsubscript{2} HoBL as well as a twisted HTL structure. Figure \ref{Figure_1}(b) displays the optical image of the stacked VdWHS, highlighting \textsuperscript{2}WSe\textsubscript{2}, \textsuperscript{1}WSe\textsubscript{2} and MoSe\textsubscript{2} regions in black, red, and blue, respectively. The twist angles in the HBL and HoBL regions, determined by polarization-resolved second-harmonic generation (SHG) measurements at RT \cite{Malard2013}, are (59$\pm$ 1)$^{\circ}$ and (54$\pm$1)$^{\circ}$ respectively, indicating that the HTL system is in a 2H-stacked configuration. Detailed information about the SHG measurements is given in the Supplementary Note (SN) 1. To investigate the distinct optical properties of the different VdWHS regions in the fabricated sample, we performed temperature-dependent PL and  valley polarization measurements, along with time-resolved PL measurements as discussed in the following.

\subsection{Room Temperature Interlayer Exctions in TMD Heterostructures}
The application of IXs in nanophotonic devices relies on their stability at RT. To observe IX emission at RT, the twist angles of the HS should be close to 60$^{\circ}$ (H-type) or 0$^{\circ}$ (R-type). In these configurations, the atomic alignment exhibits a minimal lattice mismatch and a reduced interlayer distance, which promotes efficient charge transfer efficiency between layers and increases the IX coupling strength \cite{Palekar2024,Jiang2021,Zhang2017}. Since the fabricated HBL heterostructures possess twist angles relatively close to the ideal value of 60$^{\circ}$, they almost meet the necessary conditions for the observation of IX emission at RT. Furthermore, the PL emission intensity is further enhanced by the broadband cDBR backside mirror. The dielectric cDBR is fabricated using high refractive index ($n$) contrast materials, namely  SiO\textsubscript{2} ($n =$ 1.470) and Si\textsubscript{3}N\textsubscript{4} ($n =$ 2.011), with a systematic varied layer thickness, which is deposited on a Si substrate as reported in Ref. \cite{Palekar2024_chirp}. We fabricated the cDBR with its resonance centered around 800 nm with a stopband exceeding 600 nm, which covers the emission range across all three regions of the VdWHS, as illustrated in Figure \ref{Figure_1}(c). Detailed information on the deposition parameters is provided in the SN 2. In the present cCBR design, we increased the number of mirror pairs for a resonance wavelength of 950 nm from 3 (in Ref. \cite{Palekar2024_chirp}) to 15, thereby enhancing specifically the IX emission observed at RT. 
The PL spectrum in Figure \ref{Figure_1}(d) shows the RT IX emission of the HBL region with backside cCBR, observed at 1.346 eV (921 nm). The HTL region has pronounced emission at RT, too, as presented in Figure \ref{Figure_1}(f). The HTL PL is centered and at 1.327 eV (934 nm). The corresponding red-shift of about 13 nm compared to the HBL emission, indicates interlayer hybridization in the trilayer system \cite{Palekar2024_trilayer}. Comparative RT studies of HBL and HTL regions with their monolayer counterparts are shown in Figure S2. Furthermore, the RT PL spectrum of the twisted WSe\textsubscript{2} bilayer shown in Figure \ref{Figure_1}(e) displays a bright exciton emission at 1.646 eV (753 nm), originating from the direct K-K valley transition \cite{wang2024experimental}. In addition, two lower energy emission peaks are observed at 1.5 eV (826 nm) and 1.54 eV (805 nm). These red shifted exciton peaks can be attributed to intervalley excitons involving  Q-K, Q-$\Gamma$, or K-$\Gamma$ valleys \cite{Li2022, Barman2022, Frste2020, lindlau2018role}. To further confirm the presence of intervalley exciton emission at RT, we compared the PL emission of HoBL with their monolayer counterparts (see Figure S2 in the SN 3). This comparison indicates that the additional peaks observed in the energy range of 1.5–1.6 eV (wavelength range of 826–775 nm)  are associated with intervalley excitons, which are uniquely present in the HoBL.

If we compare the PL intensities of the three VdWHS regions, distinct differences are clearly visible. Among them, the HoBL region exhibits the highest PL intensity, than that of the HBL and HTL regions, as shown in Figure \ref{Figure_1}(e). This enhanced PL intensity arises from the combined contribution of bright, dark and intervalley excitons, where both electrons and holes occupy the identical band edges in HoBL \cite{Barman2022, lindlau2018role}. In contrast, in the HBL and HTL regions, these charge carriers are localized in different layers. Interestingly, the HBL region shows about 10 times higher PL intensity than the HTL region, which we attribute to the specific twist angle configuration, spin-orbit split states such as singlet and triplet excitons, and the differences in intra- and inter-valley transitions. A detailed discussion of these mechanisms is presented in the following sections.
Overall, the near-ideal twist angles of the fabricated VdWHS system combined with the broad spectral range enhancement of the cDBR backside mirror enable stable and bright IX emission at RT. Furthermore, the distinct interlayer and intervalley exciton features observed in the HBL, HoBL, and HTL regions, and validated through comparison with the corresponding monolayers, confirm the effectiveness of this structural and optical design approach.

\subsection{Temperature-Driven Interlayer Exciton and Valley Dynamics in the MoSe\textsubscript{2}/WSe\textsubscript{2} Heterobilayer System}

\begin{figure*}
	\centering\includegraphics[width=15cm]{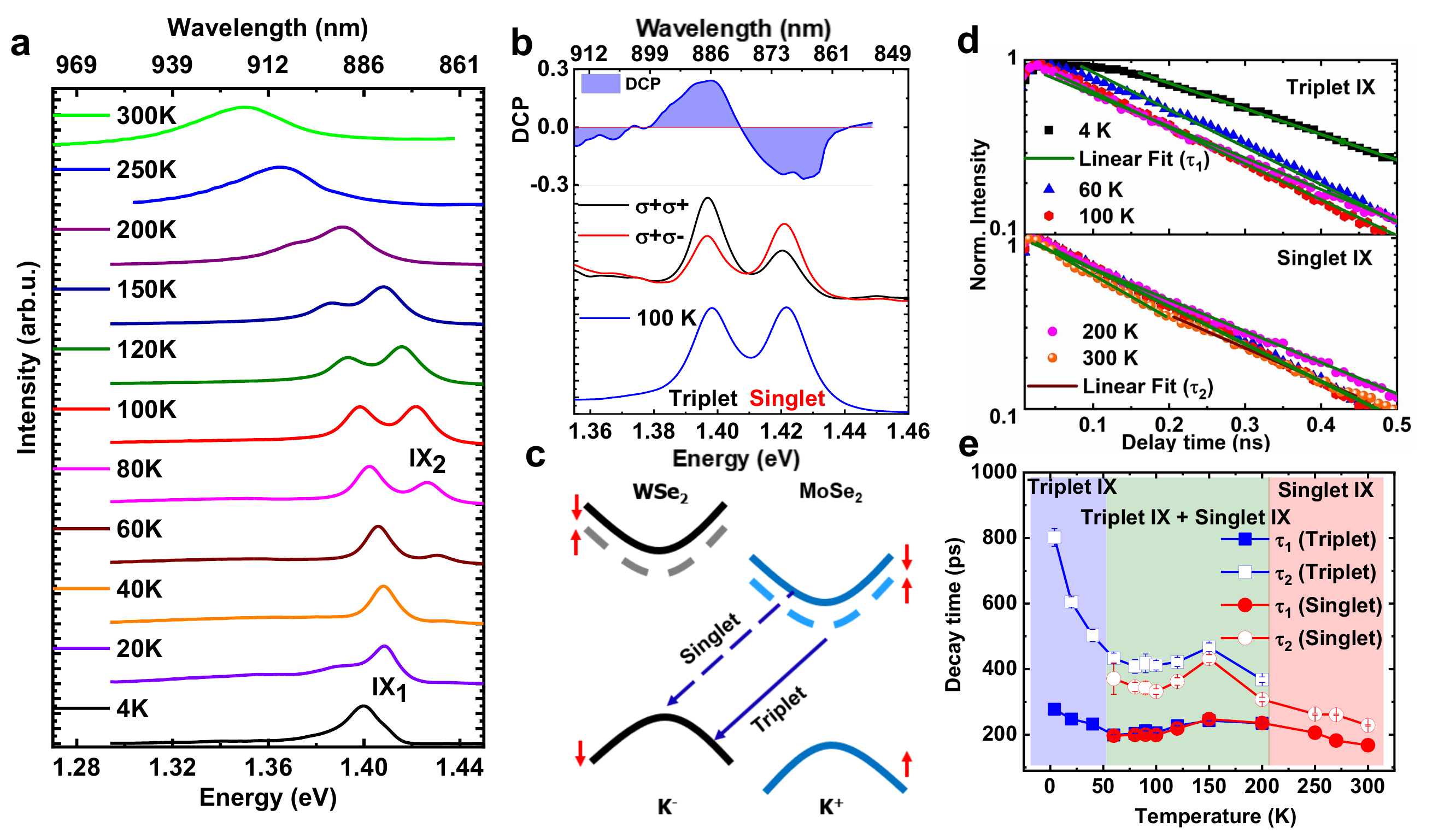} 
	\caption{\text{Excitonic emission, valley polarization, and decay dynamics of the twisted MoSe\textsubscript{2}/\textsuperscript{1}WSe\textsubscript{2} hBL.} \textbf{a} Temperature-dependent PL spectra of IX emission from 4 K to RT showing the evolution of IX emission features with increasing temperature. \textbf{b} PL spectra of IX from HBL at 100 K (bottom panel), where the triplet IX and singlet IX peaks can be well resolved. The middle panel shows polarization-resolved PL spectra at 100 K with pulsed right-circularly polarized laser excitation. ($\sigma^{+}$). The top panel represents the respective DCP. \textbf{c} Schematic illustration of the HBL band structure with a 2H stacking order, where the solid and dashed arrows correspond to triplet IX and singlet IX emission, respectively. The red arrows  represent the spin orientation of electrons and holes.  \textbf{d} Normalized PL intensity decay trace of triplet IX (top panel) and singlet IX (bottom panel)  in semi-logarithmic scale from 4~K to RT. \textbf{e} Extracted decay time components of both triplet (blue square) and singlet (red circle) IX as a function of temperature.}
	\label{Figure_2}
\end{figure*}

To gain deeper insight into the temperature stability of excitonic properties in the HBL, we performed temperature-dependent PL studies of MoSe\textsubscript{2}/WSe\textsubscript{2} HBL. These measurements were carried out over a temperature range from 4~K to RT (see Figure \ref{Figure_2}(a)). The HBL heterostructure region was excited by a picosecond pulsed laser with energy 1.72 eV (720 nm) and at a power of 50 µW. With increasing temperature, the IX (IX\textsubscript{1}) resonance energy exhibits a redshift, in agreement with previous reports \cite{meshulam2025temperature,Lou2024,Hanbicki2018}. For temperatures exceeding 40~K, an energetically higher IX (IX\textsubscript{2}) state emission is observed in Figure \ref{Figure_2}(a), in which it is seen that its contribution to the overall emission intensity progressively increases with temperature. At intermediate temperatures, for example, at 100~K, the IX emission shows a well-resolved splitting with distinct peaks at 1.398 eV (IX\textsubscript{1}) and 1.421 eV (IX\textsubscript{2}), respectively (see Figure \ref{Figure_2}(b) bottom panel). The excitation power dependence of these excitonic features measured at 100~K is shown in Fig. S3 and displays the blueshift and broadening of both IX peaks with increasing exciton density, which is consistent with previous studies \cite{Jauregui2019,Nagler2017}. The energy difference ($\Delta$E\textsubscript{IX\textsubscript{2}-IX\textsubscript{1}}) between the two peaks changes from 26 meV at 60~K to 16 meV at 200~K, indicating an electronic band splitting in the monolayers. This effect can be attributed to spin orbit coupling in monolayer TMD, which induces a small splitting of conduction band in MoSe\textsubscript{2} on the order of 20 meV \cite{Liu2013gui,Liu2015}. It has also been shown previously that the conduction band splitting gives rise to a spin triplet exciton, that is, a spin-dark exciton in MoSe\textsubscript{2} monolayers \cite{koperski2018orbital, deilmann2017dark}. However, for MoSe\textsubscript{2}/WSe\textsubscript{2} HSs with a 2H stacking symmetry, where the K\textsuperscript{+} valley of MoSe\textsubscript{2} is aligned with the K\textsuperscript{-} valley of WSe\textsubscript{2} (see level scheme in Figure \ref{Figure_2}(c), the splitting enables the formation of two configurations of the IX: spin triplet and spin singlet IX, an effect which is discussed in the following.

We performed polarization-resolved PL measurements to identify the triplet and singlet exciton contributions from these two peaks. In these studies we excited the HS with a picosecond pulsed laser of 1.72 eV ($\sigma^{+}$) right circularly polarized light and detected the PL with the same ($\sigma^{+}$) and opposite ($\sigma^{-}$) left circularly polarized light as shown in the middle panel of Figure \ref{Figure_2}(b), where the resulting PL spectra with  ($\sigma^{+}$) and ($\sigma^{-}$) are plotted at 100~K. To determine the valley polarization we define the degree of circular polarization, DCP=$\frac{I\left(\sigma^{+}\right)-I\left(\sigma^{-}\right)}{I\left(\sigma^{+}\right)+I\left(\sigma^{-}\right)}$ in which $I\left(\sigma^{+}\right)$ is the integrated intensity of the same helicity and $I\left(\sigma^{-}\right)$ is the integrated intensity of the opposite helicity. Figure \ref{Figure_2}(b), top panel, represents a positive DCP for IX\textsubscript{1} and a negative DCP for IX\textsubscript{2} at 100~K. The response of ($\sigma^{+}$) and ($\sigma^{-}$) and DCP at other temperatures is shown in Figure S4. Previous theoretical work \cite{Yu2018_spin} has shown that 2H-stacked VdWHS exhibit three different registries ($H_{h}^{x}$, $H_{h}^{h}$, $H_{h}^{m}$) with different optical selection rules for singlet and triplet states. However, only the $H_{h}^{h}$ configuration has active singlet and triplet valley polarization and exhibits negative and positive valley polarization, respectively. Therefore, here IX\textsubscript{1} and IX\textsubscript{2} can be assigned to the triplet and singlet IX that show positive and negative DCP, respectively.

Next, we performed time-resolved PL (TRPL) measurements to gain further insight into the recombination dynamics of IX emission. Figure \ref{Figure_2}(d) shows the TRPL decay curves of the IX (singlet and triplet) emission, plotted on a semi-logarithmic scale from 4~K to RT. Here, the HS region is excited with a picosecond pulsed laser of 1.72 eV (720 nm). The PL signal was measured using superconducting nanowire single-photon detectors (SNSPDs) with time resolution of 40 ps. The decay time was extracted by linear fitting for both singlet and triplet IX, considering an initial fast decay ($\tau\textsubscript{1}$) and followed by slow decay ($\tau\textsubscript{2}$) at larger delay times, as shown in Figure \ref{Figure_2}(e). Here, $\tau\textsubscript{1}$ is attributed to radiative decay of the bright excitons, which is a direct momentum transition and $\tau\textsubscript{2}$ is attributed to the decay of the phonon-assisted dark excitons \cite{Choi2021,Palummo2015}. $\tau\textsubscript{1}$ of triplet IX varies from (277 $\pm$ 1) ps at 4K to (230 $\pm$ 1) ps at 200~K and $\tau\textsubscript{2}$ varies from (800$\pm$27) ps to (367$\pm$8) ps in the same temperature range. In contrast, the singlet IX emission exhibits $\tau\textsubscript{1}$ (195 $\pm$ 1) ps at 60~K which decreases to (167 $\pm$ 4) ps at RT. Here, both IXs show significantly shorter decay time than the previously reported values \cite{Han2025, Foerg2019_opencav} and can be plausibly attributed to the combined effects of 2H type symmetry between the constituent monolayers and twist angle being close to 60$^{\circ}$. Together, these factors can lead to an increased radiative recombination rate. A comparison of the radiative decay time of the singlet and triplet IX from 60~K to 200~K reveals that their difference remains below 5 ps except for 90~K. As observed in temperature dependent PL study discussed above, $\Delta E$ between the singlet and triplet IX varies from 26 meV to 16 meV in this temperature range. As the temperature increases, the thermal energy $k_BT$ becomes comparable to $\Delta E$, enabling efficient thermal redistribution of carriers between the two states and a corresponding change of the relative PL intensities. As a result, the singlet and triplet IX emission intensities converge near thermal equilibrium, which in turn yields similar recombination dynamics at elevated temperatures. This explains the nearly identical PL decay times ($\Delta \tau$ $\leq$10 ps) observed between 60 K and 200 K suggesting that the singlet IX is in thermal equilibrium with the triplet IX \cite{Wang2020, Nagler2017NatComm}. Additionally, the reduction in radiative decay time at elevated temperatures indicates enhanced phonon-mediated non-radiative processes \cite{Choi2021}.

Overall, these investigations reveal the existence of two IX states in the 2H stacked MoSe\textsubscript{2}/WSe\textsubscript{2} HBL, arising from spin-orbit splitting in the MoSe\textsubscript{2} conduction band, which are identified as triplet and singlet IX because of their opposite valley polarizations. Their temperature-dependent redshift and broadening reflect the impact of thermal effects on their electronic band structure, and comparable recombination dynamics indicate efficient thermal redistribution of charge carriers as the thermal energy approaches their splitting. 

\begin{figure*}
    \centering\includegraphics[width=15cm]{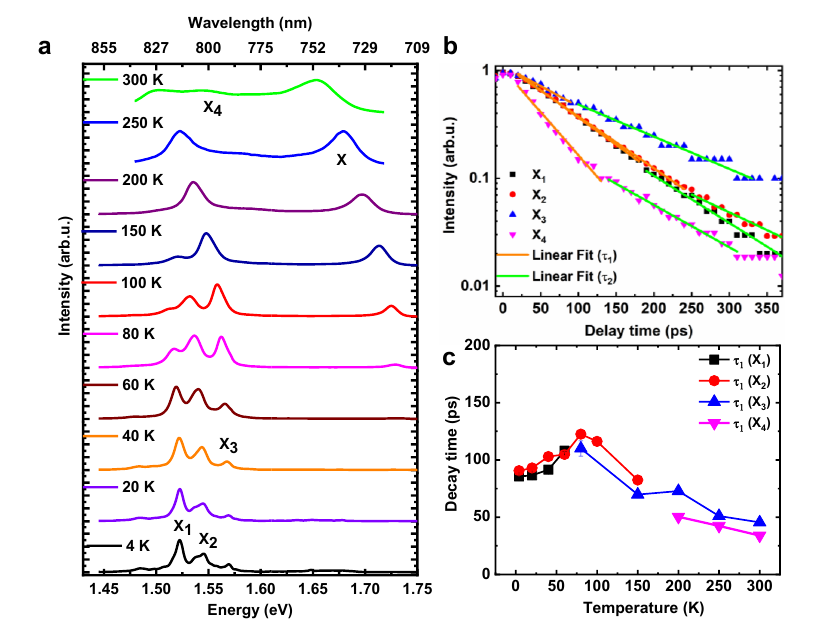}
    \caption{\text{Evolution of PL and TRPL of twisted $^{1}$WSe$_{2}$/$^{2}$WSe$_{2}$ HoBL.} 
    \textbf{a} PL spectra of the HoBL measured from 4 K to RT, showing the temperature-dependent evolution of X$_{1}$, X$_{2}$, and X$_{3}$ emission with X$_{3}$, X$_{4}$, and the intralayer exciton of WSe$_{2}$. \textbf{b} Semi-logarithmic plot of normalized PL intensity decay traces of four phonon-assisted intervalley excitons X$_1$ to X$_4$. The signals were measured at the characteristic peak temperatures: X$_{1}$ (black), X$_{2}$ (red) at 4 K, X$_{3}$ (blue) at 80 K, and X$_{4}$ (magenta) at 200 K. The characteristic decay times were extracted by linear fits. \textbf{c} Radiative decay time ($\tau_{1}$) of the HoBL obtained from linear fitting of the PL decay in the semi-logarithmic plot as a function of temperature.}
    \label{Figure_3}
\end{figure*}

\subsection{Thermal Evolution of Intervalley Exciton Response in the WSe\textsubscript{2}/WSe\textsubscript{2} Homobilayer System} 

Moving on to the three-layer system, we discuss the temperature-dependent optical response of the \textsuperscript{1}WSe\textsubscript{2}/\textsuperscript{2}WSe\textsubscript{2} HoBL in the following. Figure \ref{Figure_3}(a) displays the PL response of the HoBL region from 4~K to RT where it was excited with 660 nm CW laser at a constant power of 50 µW. At 4~K , we observe three excitonic peaks X\textsubscript{1}, X\textsubscript{2} and X\textsubscript{3} at 1.523 eV (814 nm), 1.546 eV (802 nm), and 1.569 eV (790 nm), respectively, and another peak X\textsubscript{4} appears above 200~K at 1.59 eV (779 nm). Above 80 K, the neutral WSe\textsubscript{2} exciton at 1.728 eV (717 nm) emerges. All exciton resonances of the HoBL region redshift with increasing temperature and blueshift with increasing power (see Figure S5), consistent with mechanisms discussed above for the HBL system. The four PL peaks, observed in the energy range of 1.5 and 1.6 eV, agree with the previous studies on twisted WSe\textsubscript{2} HoBL. In literature, these peaks have been interpreted differently. Some studies attribute them to moiré IX emission \cite{Wu2024,Wu2022}, whereas others assign them to phonon-assisted interlayer and intralayer intervalley excitons \cite{Huang2022, Li2022, Merkl2020, Frste2020, Wickramaratne2014}. According to Lindlau et al. \cite{lindlau2018role}, optical and acoustic phonon sidebands of QK and Q$\Gamma$ intervalley excitons dominate the spectral range from 1.57 eV-1.60 eV. In bilayer WSe\textsubscript{2}, the Q-$\Gamma$ exciton is redshifted by up to 126 meV, while the Q-K exciton is redshifted by up to 95 meV with respect to K-K exciton. Similarly, Barman et al. \cite{Barman2022} reported that the K-$\Gamma$ intervalley exciton in 2H-stacked WSe\textsubscript{2} is redshifted by 45-100 meV relative to K-K exciton.
It is noteworthy that moiré excitons reported in earlier studies are generally absent at temperatures exceeding 100 K. In contrast, X\textsubscript{3} and X\textsubscript{4} peaks in our measurements remain visible up to RT. Based on their energy position and thermal robustness, we therefore assign these peaks to intervalley excitons. A polarization dependent PL study reveals a very low DCP as displayed in Figure S7 in the SI, which is in agreement with the above reports where the valley polarization is negligible at zero magnetic field but can be enhanced by applying an external magnetic field.

To obtain more information about the nature and underlying physics of the observed emission peaks, we performed TRPL studies from 4~K to RT using the same method as for the HBL region, illustrated in Figure \ref{Figure_3}(b) and Figure S7. Figure \ref{Figure_3}(c) shows the extracted radiative decay constants ($\tau\textsubscript{1}$) of X\textsubscript{1} which varies from (85 $\pm$ 1) ps at 4~K to (108 $\pm$ 1) ps at 60~K, while X\textsubscript{2} changes from (90 $\pm$ 1) ps at 4~K to (82 $\pm$ 1) ps at 150~K.  The decay time of X\textsubscript{3} decreases from (109 $\pm$ 6) ps at 80~K to (45 $\pm$ 2) ps at RT, and that of X\textsubscript{4} reduces (50 $\pm$ 1) ps at 150~K to (33 $\pm$ 1) ps at RT. The temperature-dependent nonradiative decay times, represented by $\tau\textsubscript{2}$, of those four peaks are plotted in Figure S5. The fast decay times of these excitons with values in the range of 33 ps to 120 ps  are much longer than the reported decay time of 4 ps \cite{Wang2014, Palummo2015} for the intravalley K-K exciton. This results also indicate that the four peaks originate from phonon-assisted intervalley excitons.

\subsection{Temperature Dependence of Optical, Valley, and Exciton Dynamics in the MoSe\textsubscript{2}/WSe\textsubscript{2}/WSe\textsubscript{2} Heterotrilayer System} 

\begin{figure*}
	\centering\includegraphics[width=15cm]{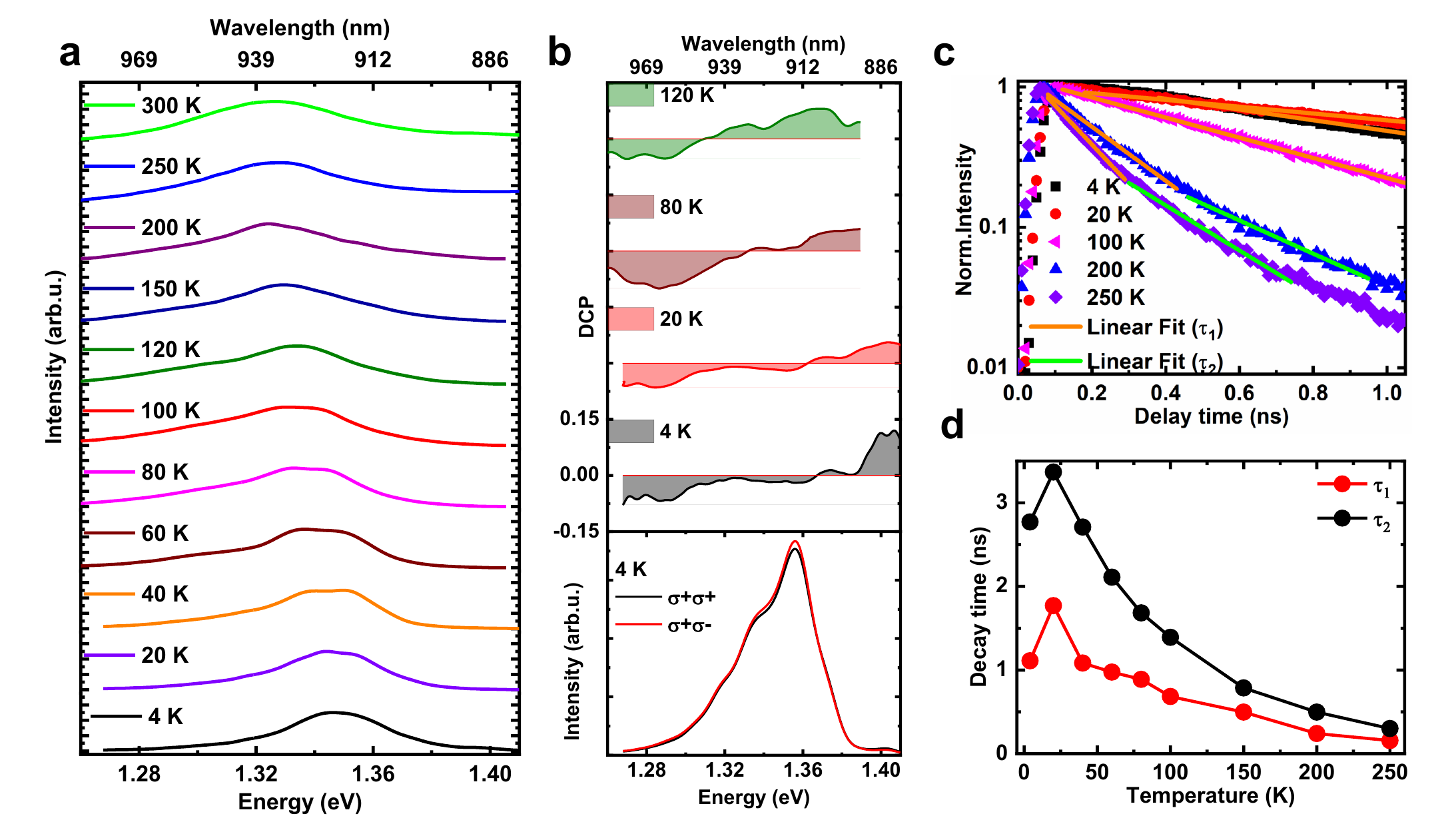} 
	\caption{\text{Emission propoerties and decay dynamics of the twisted MoSe\textsubscript{2}/\textsuperscript{1}WSe\textsubscript{2}/\textsuperscript{2}WSe\textsubscript{2} heterotrilayer system.} \textbf{a} PL spectra of HTL from 4~K to RT plotted with vertical stacking. \textbf{b} (bottom panel) Polarization resolved PL spectra under pulsed laser excitation of right circularly polarized light ($\sigma^{+}$) at 4~K. (top panel) Extracted degree of circularly polarization with respect to temperature of the HTL region. \textbf{d} Temperature-dependent TRPL decay traces in  semi-logarithmic presentation. \textbf{e)} Extracted temperature-dependent fast ($\tau\textsubscript{1}$) and slow ($\tau\textsubscript{2}$) decay constants of HTL emission. }
	\label{Figure_4}
\end{figure*}

We now turn to the heterotrilayer system, where the twisted monolayer WSe\textsubscript{2} on top of the HBL modifies crucially the exciton dynamics.  To probe optical properties of the HTL, we excited the HTL region with a pulsed laser of 1.72 eV energy at a constant power of 50 µW. First, Figure \ref{Figure_4}(a) represents the temperature-dependent PL spectra of the HTL. The emission appears at 1.347 eV at 4~K and undergoes a continued redshift with increasing temperature, reaching 1.324 eV at RT. At 4~K, the excitation power-dependent measurements reveal that the exciton resonance emission exhibits blueshfit and spectral broadening with increasing exciton density, as can be seen in Figure S8. This temperature- and power-dependent behavior of the HTL can be attributed to the same fundamental mechanisms discussed above for the HBL. The broad PL emission of the HTL with a 2H registry is in agreement with previous observations \cite{Forg2021NatComm,Palekar2024_trilayer} which suggests that this spectral feature does not correspond to a single excitonic transition, but is rather a combination of direct K-K and indirect K-Q and $\Gamma$-Q transitions of IXs. As observed in WSe\textsubscript{2} twisted bilayers, the introduction of the second monolayer creates indirect transitions between the two constituents. Similarly, when the WSe\textsubscript{2} monolayer is stacked on the HBL, it modifies the Q and $\Gamma$ valleys of the energy band \cite{Palekar2024_trilayer}.  As a result, this allows one to form IXs with more indirect momentum (K-Q, $\Gamma$-Q) compared to direct (K-K) transitions. Under the same excitation power, the hybridized exciton from HoBL is strongly quenched compared to the IX emission in the HTL region (see Figure S9 in the SI). This behavior arises because the IX resonances lie at lower energies compared to the intervalley excitons. The pronounced quenching further indicates an ultrafast and highly efficient charge transfer between MoSe\textsubscript{2} and WSe\textsubscript{2} driven by the modified energy band alignment.

To gain deeper insights into the emission properties of the HTL IX, we performed polarization-resolved PL studies using the same excitation and detection scheme as used in the HBL case. The results of these measurements are shown in Figure \ref{Figure_4}(b). The lower DCP values of HTL  approximately 0.02 compared to 0.3 for the HBL system at 4~K which is reduction by factor of 15 in DCP values indicating that the HTL shows less sensitivity to valley polarization. We attribute this observation to the fact that several transitions contributing to the IX are inactive under valley polarization, consistent with our findings for HoBL. At 4~K, the DCP is negative in the spectral range from 1.26 eV to 1.37 eV, whereas a positive DCP observes at the higher energy side of the emission peak. The positive DCP is attributed to the triplet exciton emission of the K-K valley between the \textsuperscript{1}WSe\textsubscript{2} and the MoSe\textsubscript{2} monolayers, in agreement with our observations discussed above for the HBL system. In contrast, the broadened spectral region of negative DCP further suggests that multiple singlet electronic transitions are the dominant contributors to the PL emission, which have lower energy than the direct K-K transitions between \textsuperscript{1}WSe\textsubscript{2} and MoSe\textsubscript{2}. The central singlet IX emission features are attributed to the exciton transition between \textsuperscript{2}WSe\textsubscript{2} and MoSe\textsubscript{2} as proposed by theory in Ref. \cite{Palekar2024_trilayer}. Furthermore, the lower energy singlet transitions accumulate to be momentum-indirect transitions like K-Q, $\Gamma$-Q, as observed in other reported HTL systems \cite{Forg2021NatComm}. Figure \ref{Figure_4}(b) also shows that increasing the temperature results in strongly increased negative DCP which is attributed to the dominance of intervalley singlet IX emission over triplet IX emission in the HTL region (see Figure S10 in the SI). Notably, the singlet dominance observed in the HTL region suggests that the addition of an extra WSe\textsubscript{2} monolayer to the HBL substantially modifies the excitonic wavefunctions, thereby opening distinct exciton emission channels via stacking induced bandgap engineering.

Following the valley polarization analysis, we performed TRPL studies from 4~K to 250~K to obtain insight into the relaxation dynamics of the HTL IX. Figure \ref{Figure_4}(d) shows a selection of recorded PL decay traces of the HTL IX for five temperatures form 4 K to 250 K. The experimental data were fitted with linear function to analyze the decay dynamics. The fast decay component ($\tau\textsubscript{1}$)  was extracted from the first linear fit region, followed by the extraction of the slow decay component ($\tau\textsubscript{2}$) from the second linear fitting region. The results of this data analysis are plotted in Figure \ref{Figure_4}(e) in which the black (red) data points represent the determined decay time of the slow (fast) component. At 4~K (see black curve in \ref{Figure_4}(d)) the HTL IX exhibits $\tau\textsubscript{1}$ and $\tau\textsubscript{2}$ of (1.11 $\pm$ 0.01) ns and (2.77 $\pm$ 0.15) ns, respectively. Interestingly, both decay times initially increase with temperature, reaching a maximum at 20 K with values of $\tau_1$ = (1.77 $\pm$ 0.10) ns and  $\tau_2 =$ (3.36 $\pm$ 0.15) ns, respectively. Then, a decrease in both decay times occurs at higher temperatures and  $\tau_1 = $ (0.15$\pm$0.01) ns and $\tau_2 = $ (0.26$\pm$0.01) ns are reached at 250~K. The longer decay time at lower temperatures strongly emphasizes the indirect valley nature of the HTL IX. Meanwhile, enhancement of the phonon-assisted intervalley transition and nonradiative decay channels at higher temperatures leads to a rapid reduction in decay time \cite{Miller2017,Rivera2015}.

\section{Discussion} \label{Discussion}

Having discussed the three VdWHS systems individually, we now compare the optical responses of the HTL and HBL systems to develop a comprehensive understanding of the impact of the stacking configuration. 

\begin{figure*}
	\centering\includegraphics[width=15cm]{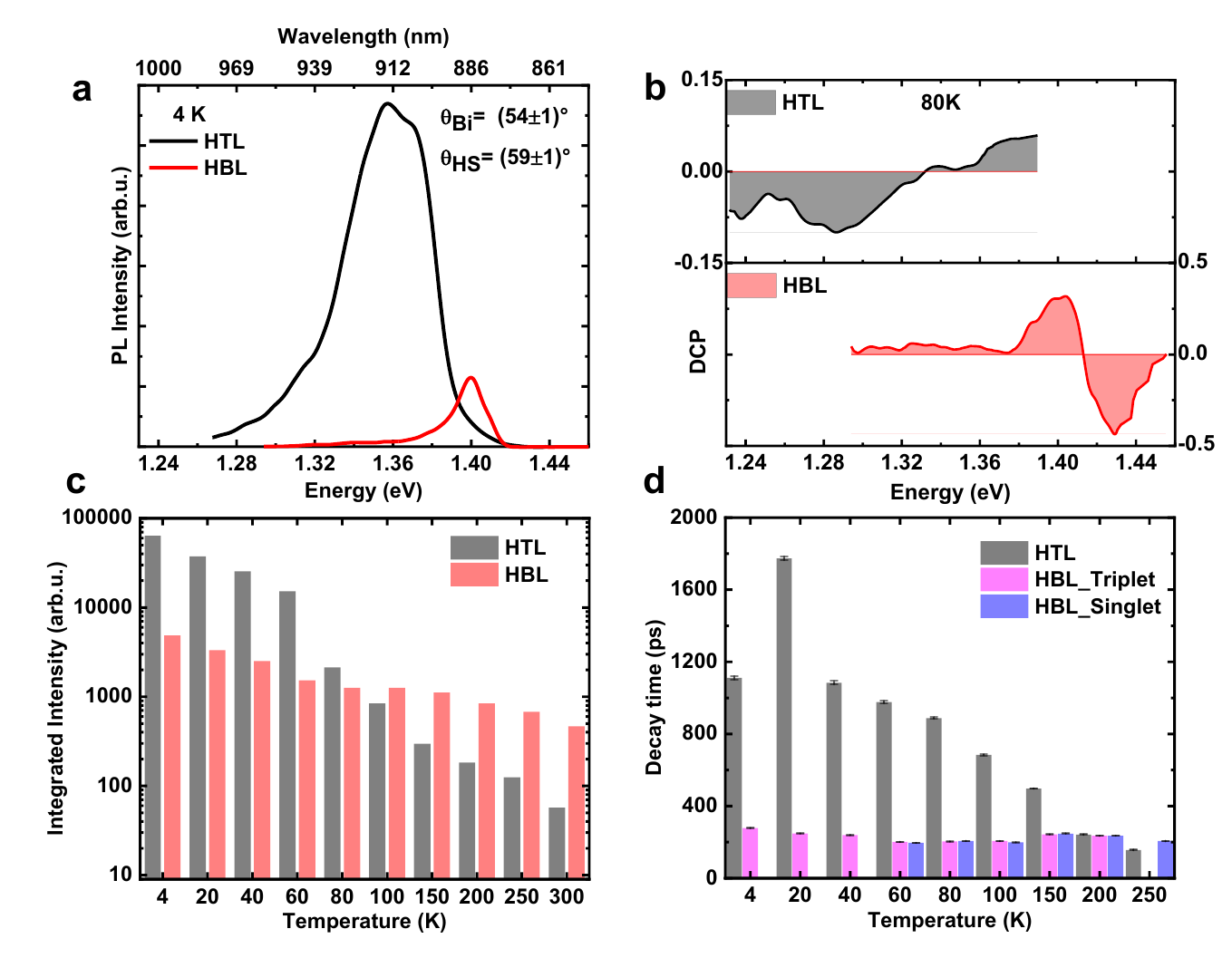} 
	\caption{\text{Comparative study of optical and valleytronic properties of IX in twisted HBL and HTL.} \textbf{a} PL spectra of both HBL (red trace) and HTL (black trace) highlighting the difference in emission features at 4~K.  \textbf{b} Calculated  DCP of both HBL and HTL obtained from right ($\sigma^{+}$) and left ($\sigma^{+}$) circular polarization upon the pulsed laser excitation with ($\sigma^{+}$) polarization at 80~K. Top panel: DCP of HTL region with the range from -0.2 to 0.2. Bottom panel: DCP of the HBL region. \textbf{c} Comparative column plot of the temperature-dependent PL intensity of HBL and HTL from 4~K to RT. \textbf{d} TRPL traces recorded over the same temperature range comparing the radiative decay times of the HBL and HTL systems.}
	\label{Figure_5}
\end{figure*}

We first compare the emission intensity of the HTL and HBL systems at 4~K in Figure \ref{Figure_5}(a) which shows a significant increase in PL intensity from the HTL region (black curve) compared to the PL intensity from the HBL region (red curve) under identical excitation conditions. Comparing the integrated PL intensity shows that adding the third monolayer to the HTL configuration increases IX intensity by a factor of ten relative to the HBL system. The observed emission enhancement is larger than in the 2H-stacking symmetry reported in Ref. \cite{Palekar2024_trilayer}, predominantly
 due to low twist angle between the constituent layers. For additional information, we refer to Figure S11 in the SI, which illustrates the enhancement of HTL compared to HBL at different spots on the HS. We attribute the increase in IX intensity to three effects, namely a) additional optical absorption of the laser light in the \textsuperscript{2}WSe\textsubscript{2} layer, b) efficient charge transfer between WSe\textsubscript{2} layers, and c) appearance of momentum indirect channels. This interpretation was confirmed by PLE and reflectance measurements in Ref. \cite{Palekar2024_trilayer}. Figure S10 contains spectral information of both HTL IX and WSe\textsubscript{2} HoBL exciton at 4~K, showing the highly quenched emission from the X\textsubscript{HoBL} compared to the HTL IX, which indicates ultra-fast charge transfer between MoSe\textsubscript{2} and WSe\textsubscript{2} layers. The redshift in the HTL emission, which is about 41(±1) meV compared to the HBL, also supports band modifications and the contribution of indirect momentum transitions in the HTL due to the addition of WSe\textsubscript{2} as indirect transitions such as K-Q and $\Gamma$-Q valley transitions are at lower energy compared to the direct K-K valley transition \cite{Palekar2024_trilayer,Forg2021NatComm}. 

Additional insight is obtained by a comparison of the valley polarization between HBL and HTL. From the individual analyses discussed in the Results section, it is evident that at 4~K, the IX of both HTL and HBL is dominated by singlet and triplet IX states, respectively. However, with increasing temperature, the HBL emission includes contributions from both singlet and triplet IX because of spin orbit coupling, while for the HTL the emission is dominated by singlet IX states. The results of polarization dependent PL studies of HBL and HTL regions at 80~K are shown in Figure \ref{Figure_5}(b). The HBL exhibits positive and negative DCP in the 1.38-1.44 eV spectral range, while the HTL shows pronounced negative DCP across the wider 1.28-1.37 eV range. These DCP signatures reveal clearly that the PL resonance in the HBL originates only from direct K-K valley excitons with different spins, while in the HTL system the addition of a \textsuperscript{2}WSe\textsubscript{2} monolayer leads to band hybridization, giving rise to both direct and indirect valley excitons with same spin as observed in the valley polarization of HTL.

Next, we compare the temperature-dependent optical response of the HBL and HTL systems. Figure \ref{Figure_5}(c) shows the corresponding evolution of integrated PL intensity from 4~K to RT. As discussed above, at 4~K the HTL exhibits a 10-fold higher intensity compared to the HBL. However, this difference decreases significantly with increasing temperature, and beyond 80~K the PL intensity of the HBL surpasses the intensity of the HTL, as the HTL intensity decreases stronger with temperature. This significantly different temperature-dependent behavior of HTL and HBL, with more stable emission intensity of the latter, can be attributed to different aspects. The first one is related to the twist angle between the monolayers. The HBL exhibits a twist angle of (59$\pm$ 1)$^{\circ}$, whereas, in the HTL the \textsuperscript{2}WSe\textsubscript{2} forms a twist angle of (54$\pm$ 1)$^{\circ}$ with the underlying HBL. Previous studies have shown that increasing the twist angle leads to greater interlayer separation and lattice mismatch. This, in turn, alters the band alignment and governs the coupling strength within the VdWHS \cite{Palekar2024,Sebait2023}. Consequently, the twist angle dependent interlayer coupling affects the temperature stability of emission, with HBL maintaining higher intensity at elevated temperatures, while the HTL exhibits a more pronounced decrease in intensity due to weaker coupling and increased non-radiative process.  The second factor causing the different temperature dependence of the HTL and HBL systems is the involvement of momentum indirect exciton in the HTL. As discussed above, indirect momentum transitions contribute significantly to the PL of the HTL. These excitons have lower oscillator strength and require phonon assistance to initiate radiative recombination \cite{Barre2022Science}. With increasing temperature, the enhanced phonon population promotes redistribution of this intervalley IX into dark or nonradiative states and increases the dissociation of free carriers \cite{sarkar2025exciton,perea2021phonon}. This process suppresses the intervalley IX emission, which forms a major part of the HTL PL signal. Therefore, the PL intensity decreases more rapidly with temperature than for the HBL system, whose PL is dominated by the direct IX. The last and most important aspect is the thermal population of the electronic states contributing to PL emission. As observed in HBL, beyond 60~K higher energy singlet states emerges and becomes thermally populated with increasing temperature. Since the singlet IX possesses a larger oscillator strength than the triplet IX \cite{Lou2024}, this thermally populated singlet state compensates for the reduction in quantum efficiency of the triplet state caused by non-radiative decay processes in the HBL system. In contrast, in the HTL system, the thermally populated states are either singlet momentum-indirect or triplet momentum-direct states, both exhibiting low oscillator strength, which accounts for the strong temperature-dependent decrease of the HTL PL. However, this system also has a prolonged singlet state, ensuring emission at higher temperature, although with reduced intensity, because it is still affected by the non-radiative process  at elevated temperatures.

The discussed aspects are further supported by the temperature-dependent decay behavior of HTL and HBL systems as shown in Figure \ref{Figure_5}(d). At 20~K, the HTL has a decay time nearly 7 times longer than that of HBL. This large difference indicates the band structure modifications in the HTL compared to HBL, which increase the radiative decay time of the HTL. However, the decay time difference between HTL and HBL (singlet and triplet) decreases continuously with temperature, and it is inverted at 250 K. The decay time between 4 K and 250 K, $\frac{\tau_{250\,\mathrm{K}} - \tau_{4\,\mathrm{K}}}{\tau_{4\,\mathrm{K}}} \times 100\%$
 shows a reduction of about $\sim 86\%$ for HTL, while the HBL, considering both triplet and singlet contributions, exhibits a significantly lower reduction of approximately $\sim 25\%$. The stronger reduction observed in the HTL highlights its pronounced dependence on indirect valley transitions from singlet IX states, reflecting a distinct temperature sensitivity compared to that of the HBL.

In summary, we observed striking, temperature-dependent differences in the optical behavior of 2H-stacked HTL and HBL systems from 4 K to 300 K. The twisted WSe\textsubscript{2} on HBL enables the HTL to achieve over a tenfold enhancement in PL intensity and a sevenfold increase in excitonic decay at low temperatures. The HTL and HBL respond differently to temperature variations, as reflected in their temperature-dependent valley polarization, confirming their unique valley splitting and excitonic responses with temperature. This demonstrates the significant potential of VdWHS engineering in shaping and managing the optical properties of resulting multilayer systems.  

\section{Conclusion} \label{Conclusion}

By engineering the layer design and twist angles, we achieved RT IX emission in a VdWHS system, revealing distinct and previously unexplored temperature-dependent optical behavior. A trilayer heterostructure with a 2H-type stacking configuration was fabricated, enabling the formation of well-defined HBL, HoBL, and HTL regions within a single VdWHS system. The TMD heterostructures were integrated onto a broadband (600 nm) chirped distributed Bragg reflector composed of SiO\textsubscript{2}/Si\textsubscript{3}N\textsubscript{4} mirror pairs to enhance PL intensity. The cDBR-enhanced luminescence facilitated optical studies at elevated temperatures and represents a significant step toward practical excitonic and optoelectronic devices based on VdWHS.

We identified, for the first time, temperature-dependent variations in singlet and triplet IX decay times, where the small energy splitting of 20 meV supports thermal equilibrium at elevated temperatures and enables comparable recombination dynamics. In the HoBL region, PL, valley polarization, and decay dynamics were analyzed, although a complete understanding of valley-specific emission remains to be developed. For the HTL, we report the first systematic study of temperature-dependent PL, polarization, and decay times, revealing a red-shifted emission energy, negative degree of circular polarization, and increased exciton decay time relative to the HBL. These features indicate pronounced band modulation and the presence of interlayer and intervalley transitions. Notably, the HTL exhibits a tenfold enhancement in PL intensity and a sevenfold increase in radiative decay time at 4 K compared to the HBL, while also showing stronger temperature sensitivity. This highlights the crucial influence of twist angle, intervalley exciton processes, and thermally populated states. Overall, our findings establish a clear strategy for engineering exciton dynamics through stacking orientation and Bragg resonators, laying the foundation for next-generation optoelectronic devices \cite{Wang2020, Ross2017}, quantum computing platforms \cite{Jiang2021, Sohoni2020}, and energy-harvesting applications \cite{Lin2022, Jiang2021}.

\section{Acknowledgments}

B.S. Sahoo, S. Machchhar, C.C. Palekar and S. Reitzenstein acknowledge financial support by the Deutsche Forschungsgemeinschaft (DFG) within the Priority Program SPP 2244 “2DMP” project Re2974/26-1 (ID 443416027) and by the Berlin Senate via Berlin Quantum.

S.A.Tongay acknowledges primary support from DOE-SC0020653 (optical studies on quantum crystals). Partial support comes from NSF CBET 2330110 (environmental stability tests) and NSF CMMI 2129412 (manufacturing). S.A.Tongay also acknowledges partial support from Applied Materials Inc. and Lawrence Semiconductor Labs for deposition systems.

K.Watanabe and T.Taniguchi acknowledge support from the JSPS KAKENHI (Grant Numbers 21H05233 and 23H02052) , the CREST (JPMJCR24A5), JST and World Premier International Research Center Initiative (WPI), MEXT, Japan.


\section{Methods} \label{Methods}
\subsection{TMD Sample Preparation}
To investigate the exciton and valley dynamics of HBL, HTL, and HoBL structures based on MoSe\textsubscript{2} and WSe\textsubscript{2} monolayer, a polymer-assisted dry transfer technique using PC/PDMS stamping was employed for deterministic assembly of a VdWHS system. An 8 wt\% polycarbonate (PC) solution was prepared by dissolving PC in chloroform overnight, following procedures adapted from established polymer-assisted transfer methods \cite{Kim2016, Pizzocchero2016, Zomer2014}. To form the transfer film, the PC solution was first drop-cast onto a clean glass slide and allowed to dry completely at RT for five minutes, yielding a uniform solid PC layer. This freestanding PC film was then carefully lifted from the glass slide and laminated onto a PDMS stamp mounted on a second glass slide. Monolayers of MoSe\textsubscript{2}, WSe\textsubscript{2}, and few layer hBN were mechanically exfoliated onto SiO\textsubscript{2}/Si substrates (285 nm) using standard micromechanical cleavage techniques \cite{Novoselov2004}. For the pick-up step, the PC/PDMS stamp was aligned over the target WSe\textsubscript{2} monolayer and brought into contact at 120 °C on a temperature controlled micromanipulation stage. After thermal softening of the PC layer, the stamp was slowly retracted to lift the few layers of hBN,  \textsuperscript{2}WSe\textsubscript{2}, and \textsuperscript{1}WSe\textsubscript{2} flakes, consistent with hot pick-up approaches described previously \cite{Pizzocchero2016, Zomer2014}. The hBN/\textsuperscript{2}WSe\textsubscript{2}/\textsuperscript{1}WSe\textsubscript{2}/PC/PDMS stack was then aligned with the MoSe\textsubscript{2} monolayer and brought into contact at 120 °C to assemble the vdW HTL. To transfer the HTL onto the cDBR, the PC/PDMS/flake assembly was placed onto the cDBR substrate at 200 °C to ensure strong adhesion of the PC film. The PDMS block was peeled back gradually, leaving the sample with the PC film stack on the substrate. The sample was immersed in chloroform for 20–40 min to dissolve the PC completely, followed by rinsing in isopropanol and drying with nitrogen. This procedure produced a clean HTL covered with an hBN, and ensured layer interfaces with minimal polymer contamination.
  
\subsection{Spectroscopy Methods}

The sample was mounted in a cryogenic chamber equipped with active temperature control and piezoelectric nanopositioners. A white-light illumination source (LED) was used to visualize the sample and locate the excitation spot on a CMOS camera. The sample could be excited using multiple laser sources. Those include a wavelength-tunable picosecond pulsed laser (repetition frequency: 80 MHz) and CW lasers operating emitting at 660 nm and 532 nm. For RT monolayer identification, a CW laser at 532 nm served as the excitation source. Cryogenic and temperature-dependent measurements were performed using a closed-cycle cryostat equipped with a high-numerical-aperture microscope objective (NA = 0.81) and integrated nanopositioners for precise sample alignment. During optical measurements, the excitation beam was focused onto the sample by the low-temperature objective, which was also used to collect the resulting PL. The PL signal was then coupled into a spectrometer with a resolution of 430 pm. For the TRPL study, the spectrally filtered output from the spectrometer was fiber-coupled to a superconducting nanowire single-photon detector (SNSPD) connecting to a time-tagging module with a combined time resolution of less than 30 ps. Second-harmonic generation (SHG) measurements used to determine the twist angle were performed using the same optical setup, with both excitation and detection paths equipped with polarization-selective optics.

\section{Conflict of Interest} 
The authors declare no conflict of interest.


\section{Reference}
\bibliography{sorsamp}



\end{document}


\preprint{APS/123-QED}

\title{Robust Interlayer Exciton Interplay in Twisted van der Waals Heterotrilayer on a Broadband Bragg Reflector up to Room Temperature}

\author{Bhabani Sankar Sahoo}
\affiliation{Institut für Physik und Astronomie, Technische Universit\"at Berlin, Hardenbergstrasse 36, 10623 Berlin, Germany}

\author{Shachi Machchhar}%
\affiliation{Institut für Physik und Astronomie, Technische Universit\"at Berlin, Hardenbergstrasse 36, 10623 Berlin, Germany}

\author{Avijit Barua}%
\affiliation{Institut für Physik und Astronomie, Technische Universit\"at Berlin, Hardenbergstrasse 36, 10623 Berlin, Germany}

\author{Martin Podhorský}%
\affiliation{Institut für Physik und Astronomie, Technische Universit\"at Berlin, Hardenbergstrasse 36, 10623 Berlin, Germany}

\author{Seth Ariel Tongay}%
\affiliation{Materials Science and Engineering, School for Engineering of Matter, Transport and Energy, Arizona State University, Tempe, USA}

\author{Takashi Taniguchi}%
\affiliation{Research Center for Materials Nanoarchitectonics, National Institute for Materials Science,  1-1 Namiki, Tsukuba 305-0044, Japan}

\author{Kenji Watanabe}%
\affiliation{Research Center for Electronic and Optical Materials, National Institute for Materials Science, 1-1 Namiki, Tsukuba 305-0044, Japan}

\author{Chirag Chandrakant Palekar}%
\email{c.palekar@tu-berlin.de}
\affiliation{Institut für Physik und Astronomie, Technische Universit\"at Berlin, Hardenbergstrasse 36, 10623 Berlin, Germany}

\author{Stephan Reitzenstein}%
\affiliation{Institut für Physik und Astronomie, Technische Universit\"at Berlin, Hardenbergstrasse 36, 10623 Berlin, Germany}

\date{\today}

\maketitle

\vspace*{2em} 
\begin{center}
    \textbf{\LARGE Supplementary material}
\end{center}
\vspace{1em} 


\section{Supplementary Note 1: Twist Angle Determination by Second-Harmonic Generation} \label{Supplementary}

\begin{figure}
	\centering\includegraphics[width=15cm] {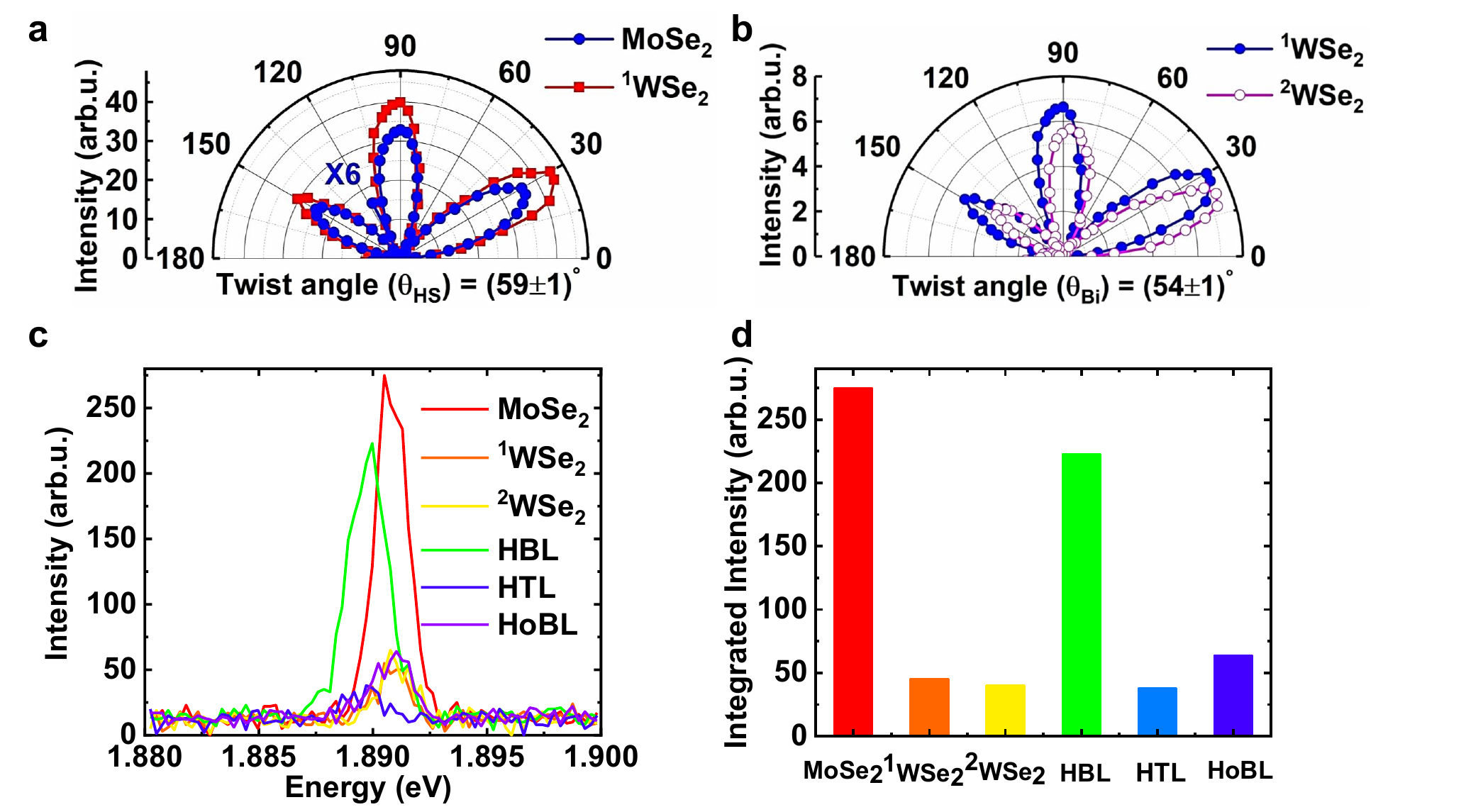} 
	\caption{\text{Twist angle measurements by SHG.} \textbf{a,b} SHG signals from the HBL and HoBL regions with different twist angles measured as a function of the excitation laser polarization angle. The polarization-resolved SHG response was used to identify the crystallographic orientations of the constituent layers. \textbf{c} The SHG intensity maxima of each monolayer. \textbf{d} Histogram representation of the SHG intensity.}
	\label{Figure_6}
\end{figure}

The twist angle between the monolayers of the van der Waals heterostructure (VdWHS) was determined using polarization-resolved second-harmonic generation (SHG) measurements. The SHG excitation was performed using a picosecond pulsed laser operating at 1313 nm, and the generated SHG signal was detected at 656 nm. The constituent monolayers of the VdWHS were illuminated with linearly polarized laser light, and the SHG intensity was recorded as a function of the incident polarization angle. Each transition metal dichalcogenide (TMD) monolayer exhibits a characteristic six-fold rotational symmetry in its SHG response. The six intensity maxima correspond to the armchair crystallographic directions, enabling precise crystallographic orientation mapping of individual MLs. By comparing the polarization-dependent SHG patterns of the monolayers within the heterobilayer (HBL) and homobilayer (HoBL), the relative crystallographic alignment—and hence the twist angle, was determined as shown in Figure \ref{Figure_6}(a) and Figure \ref{Figure_6}(b), respectively. To distinguish between stacking configurations of R-type and H-type, we performed SHG measurements on both the isolated MLs and the corresponding HS regions, as displayed in the Figure \ref{Figure_6}(c),\ref{Figure_6}(d). A distinct reduction in SHG intensity from the HBL and heterotrilayer (HTL) regions relative to the monolayers indicates 2H-type stacking, arising from destructive interference between the constituent layers, as they create an inversion symmetry \cite{Hsu2014}. Comparing the SHG response from the measured HS gives twist angle of (59$\pm$1)$^{\circ}$ for HBL and (54$\pm$1)$^{\circ}$ for HoBL, which leads to formation of 2H type symmetric stacking configuration within HTL. 

\newpage

\section{Supplementary Note 2: Design Details of the Chirped Distributed Bragg Reflector Deposition} 

The parameters listed in Table \ref{tabble_1} were used to model the reflection characteristics of the chirped distributed Bragg reflector (cDBR) using the transfer matrix method (TMM). The resulting calculated reflection spectrum, shown in Figure 1 (c), reveals a broad stopband exceeding 600 nm in width. The reflection response of the multilayer DBR stack was obtained using a modified TMM based simulation code optimized for chirped structures \cite{Wall2020}.

\label{table_1}
\begin{table}[ht]
\centering
\caption{Design parameters of the bottom chirped distributed Bragg reflector}
\begin{tabular}{l c}
\toprule
\textbf{DBR Parameters} & \textbf{Bottom DBR} \\
Materials and refractive index 
& SiO$_2$ ($n_1=1.470$), Si$_3$N$_4$ ($n_2=2.011$) \\
Layer thickness 
& $\lambda/4$ \\
Total wavelength steps 
& 9 \\
Starting wavelength $\lambda_1$ 
& 950 nm \\
Wavelength separation 
& 80 nm \\
Pair repetition for each $\lambda$ except $\lambda_1$ 
& 3 \\
Pair repetition for starting wavelength 
& 15 \\
\end{tabular}
\label{tabble_1}
\end{table}
\newpage
\begin{figure}[h]
	\centering\includegraphics[width=15cm]{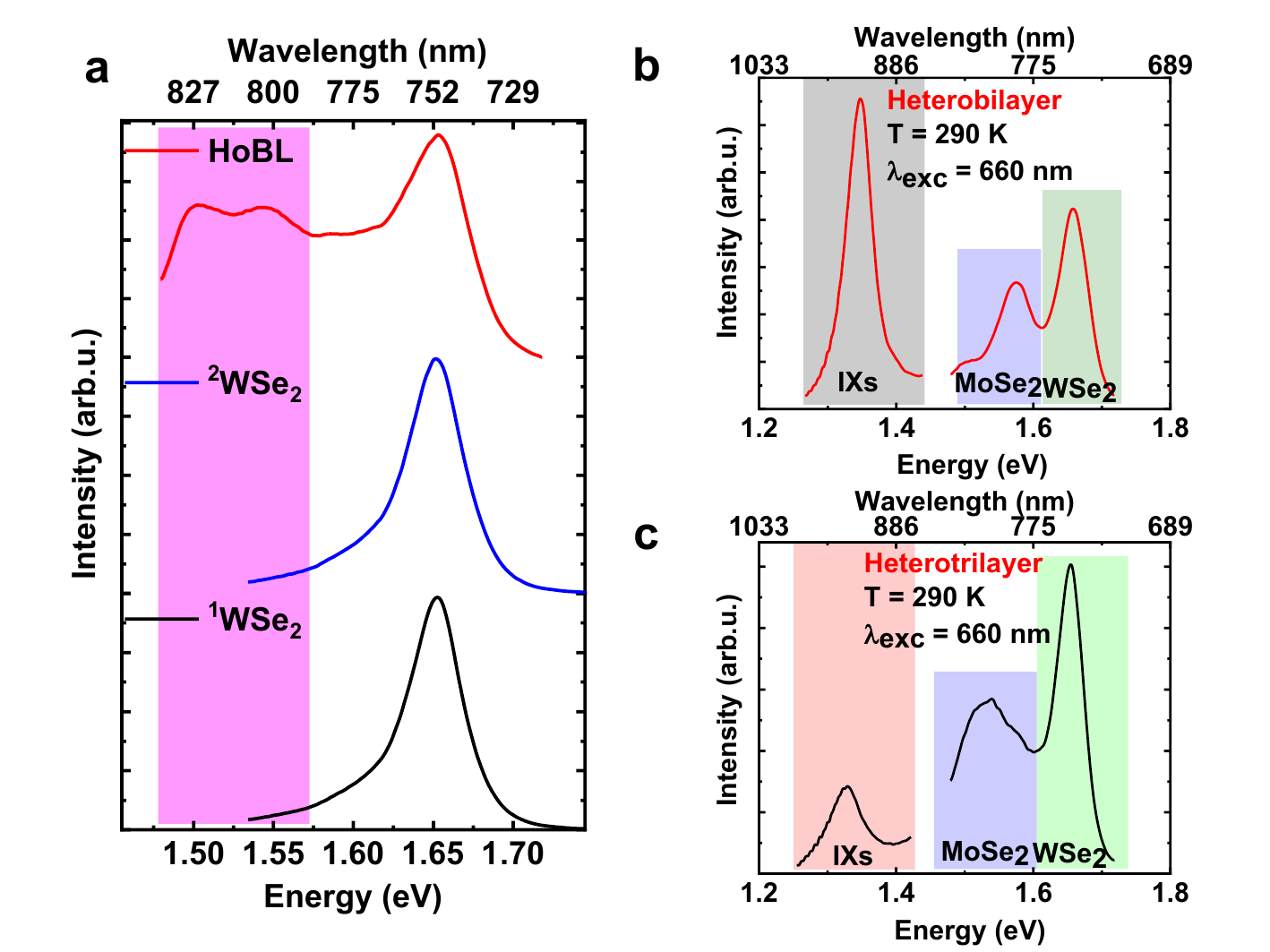} 
	\caption{\text{ RT PL emission of HoBL, HBL, HTL, and their monolayers.} \textbf{a} Top panel showing RT PL emission from HoBL region. Middle and bottom panel: RT PL of top and bottom monolayer WSe\textsubscript{2} respectively. \textbf{b} RT PL emission from HBL region which contain IX and intralayer exciton of its monolayers. \textbf{c} RT emission of IX in the HTL and its monolayer counterparts.}
	\label{Figure_7}
\end{figure}
\section{Supplementary Note 3: Room Temperature Photoluminescence of Heterostructures and their Monolayer Counterparts.}

Room temperature (RT) Photoluminescence (PL) measurements were performed on the stacked HS and its constituent monolayers using a CW laser at 660 nm. We first compare the emission from the WSe\textsubscript{2} HoBL region with that of the individual WSe\textsubscript{2} monolayers as shown in Figure \ref{Figure_7}(a). All monolayers exhibit the characteristic direct K–K intralayer exciton emission at 1.65 eV \cite{He2014, Zeng2013}. In contrast, the HoBL shows additional features at 1.50 eV and 1.55 eV that are absent in the ML spectra. These additional peaks are attributed to phonon-assisted excitonic transitions, arising from interlayer coupling within the HoBL. As displayed in Figure \ref{Figure_7}b), the HBL region exhibits a prominent IX  peak at 1.346 eV, accompanied by intralayer exciton emission originating from the constituent monolayers, specifically, the MoSe\textsubscript{2} intralayer exciton at 1.57 eV \cite{Li2021} and the WSe\textsubscript{2} intralayer exciton at 1.65 eV. Similarly, the HTL region (see Figure \ref{Figure_7}(c)) shows a distinct IXs emission at 1.327 eV. In addition to the WSe\textsubscript{2} intralayer exciton at 1.65 eV, the HTL spectrum presents a broader feature near the MoSe\textsubscript{2} intralayer exciton energy as well as contributions from interlayer excitons (IX) between the coupled WSe\textsubscript{2} HoBL. These signatures confirm the presence of multiple IX channels in HTL enabled by the trilayer stacking configuration.

\begin{figure}[h]
	\centering\includegraphics[width=15cm]{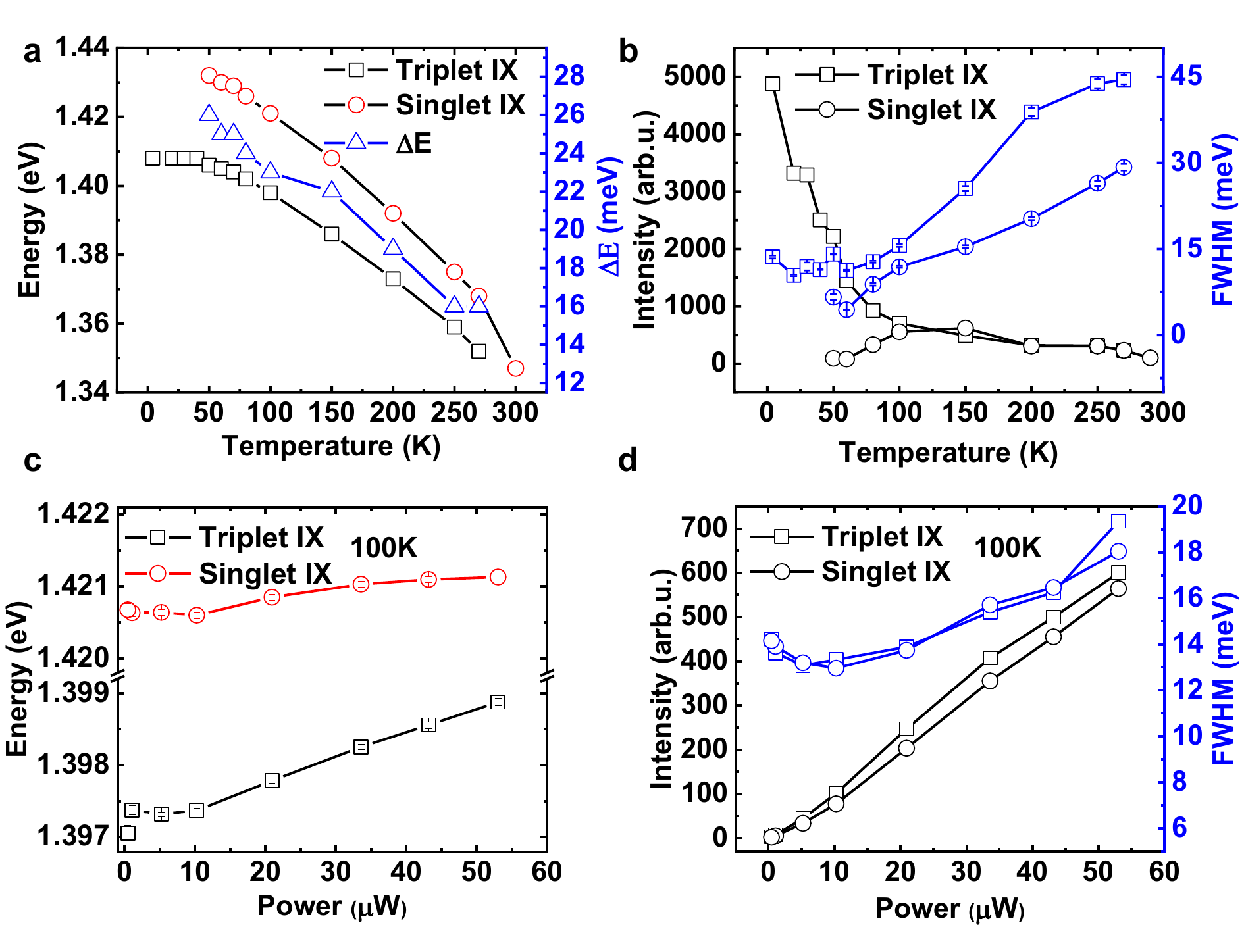} 
	\caption{\text{Temperature- and power-dependent PL characteristics of the HBL.} \textbf{a} Variation of IX emission from triplet and singlet states as a function of temperature. \textbf{b} Temperature-dependent variation of IX intensity and FWHM. \textbf{c,d} Power-dependent evolution of energy, intensity, and FWHM of triplet and singlet IX.}
	\label{Figure_8}
\end{figure}

\section{Supplementary Note 4: Temperature- and Power-dependent Photoluminescence of the HBL}

The temperature- and power-dependent PL response of the singlet and triplet IXs of the HBL was systematically investigated under pulsed excitation at 720 nm. Figure \ref{Figure_8}a) shows the variation of the IX emission energy with temperature, while Figure \ref{Figure_8}(b) depicts the corresponding changes in PL intensity and FWHM. With increasing temperature, the emission energy redshifts, accompanied by a reduction in intensity and broadening of the linewidth, indicating enhanced phonon interactions and non-radiative recombination  \cite{Rivera2015, Seyler2019Nature}. Figures \ref{Figure_8}(c, d) illustrate the power-dependent behavior, revealing the evolution of the IX emission energy, intensity, and FWHM with increasing excitation power, consistent with exciton-exciton interaction effects reported in earlier studies \cite{Unuchek2019}. These measurements provide insight into the thermal stability and exciton dynamics in the HBL under different excitation conditions.

\begin{figure}[h]
	\centering\includegraphics[width=15cm]{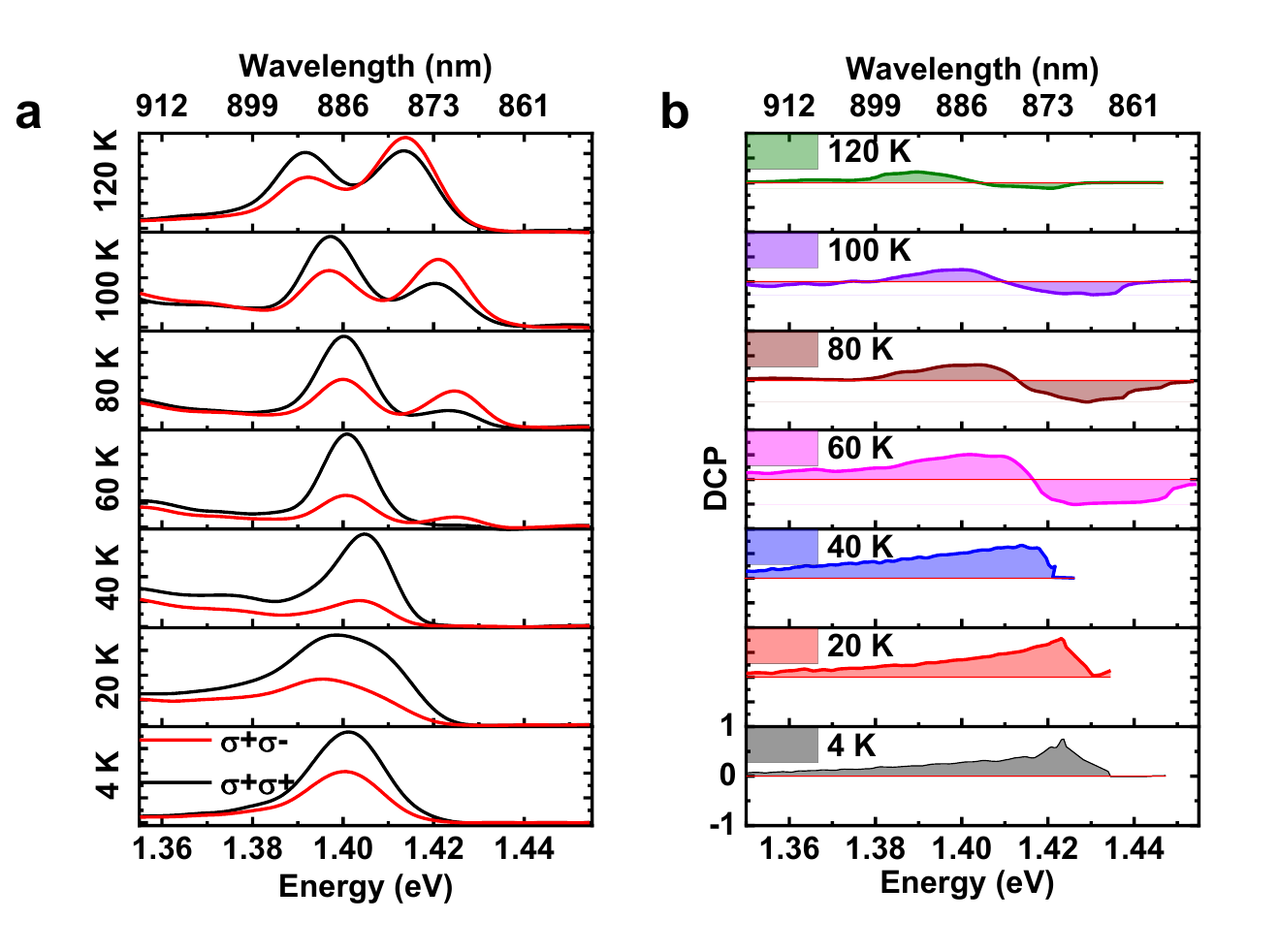} 
	\caption{\text{Temperature-dependent valley polarization of the HBL.} \textbf{a} Temperature-dependent PL emission recorded for right ($\sigma^{+}$) and left ($\sigma^{+}$) circular polarization upon the pulsed laser excitation with $\sigma^{+}$ polarization of HBL \textbf{b} Corresponding temperature dependence of degree of circular polarization (DCP) of the HBL.}
	\label{Figure_9}
\end{figure}

\section{Supplementary Note 5: Temperature-dependent Valley Polarization of HBL}

The valley polarization of the HBL was probed using a pulsed laser excitation at 720 nm with right circular polarization ($\sigma^+$). Temperature-dependent PL spectra were recorded for both co-polarized ($\sigma^+$) and cross-polarized ($\sigma^-$) emission components, as shown in Figure \ref{Figure_9}a). From these measurements, the degree of circular polarization (DCP) was calculated and plotted as a function of temperature in Figure \ref{Figure_9}b). These results reveal the temperature-dependent evolution of the singlet and triplet IXs in the HBL, providing important insights into the spin–orbit coupling in the HBL HS.
\begin{figure}[h]
	\centering\includegraphics[width=15cm]{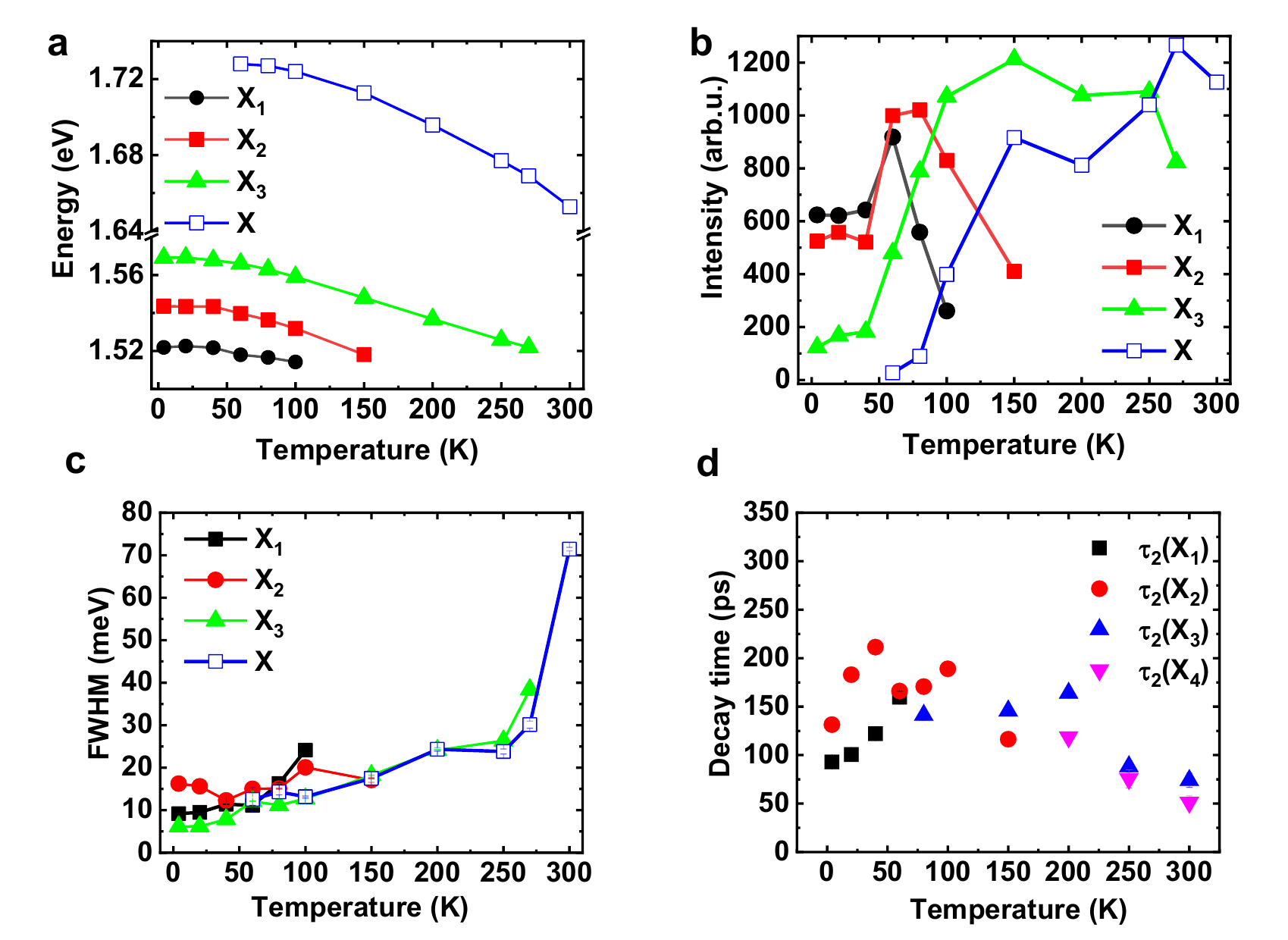} 
	\caption{\text{Temperature-dependent PL and TRPL characteristics of phonon assisted excitons in the HoBL.} \textbf{a-c} Evolution of emission energy, PL intensity, and FWHM of phonon assisted excitons and the direct intravalley exciton. \textbf{d} Temperature-dependent nonradiative decay (slow decay($t$\textsubscript{2})) times extracted from TRPL measurements.}
	\label{Figure_10}
\end{figure}

\section{Supplementary Note 6: Temperature-Dependent Photoluminescence Characteristics of the HoBL}

Figure \ref{Figure_10} shows the temperature-dependent PL and TRPL behavior of phonon-assisted intervalley (X\textsubscript{1}, X\textsubscript{2}, X\textsubscript{3}) and K-K intravalley (X) excitons in the HoBL. Panels (a–c) present the emission energy, PL intensity, and FWHM of the phonon-assisted excitons, while panel (d) displays the nonradiative decay times extracted from TRPL measurements. PL was measured with a 660 nm continuous-wave laser, and lifetimes were obtained using a 720 nm pulsed laser. These results complement the temperature- and power-dependent PL characteristics discussed for the HBL in Supplementary Note 4, highlighting the distinct exciton dynamics in the HoBL.

\section{Supplementary Note 7: Valley Polarization of the HoBL}

\begin{figure}[h]
	\centering\includegraphics[width=15cm]{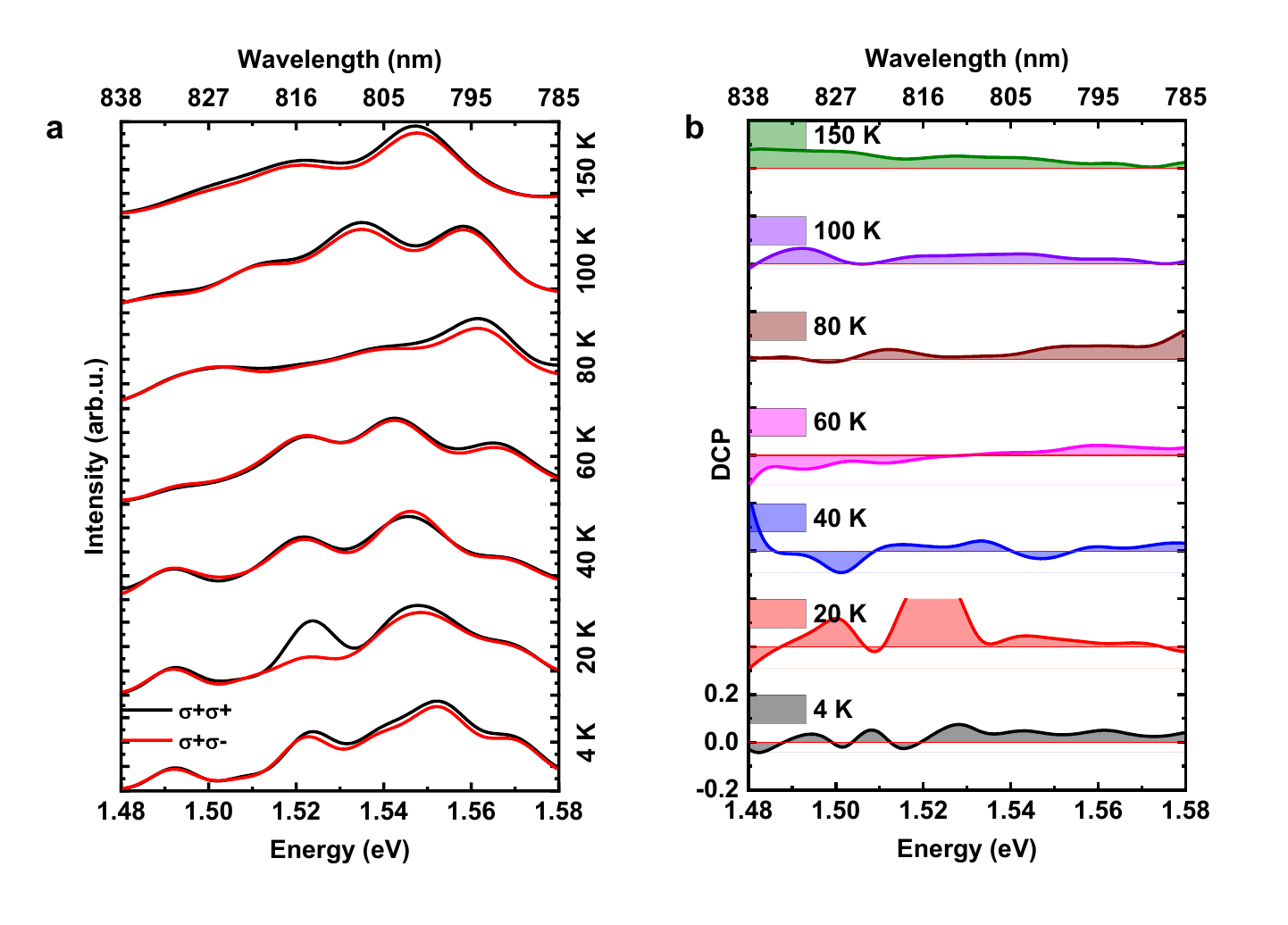} 
	\caption{\text{Temperature-dependent valley polarization of the HoBL.} \textbf{a} Temperature-dependent PL emission recorded for $\sigma^{+}$ and $\sigma^{+}$ circular polarization upon the pulsed laser excitation with $\sigma^{+}$ polarization of HoBL \textbf{b} Corresponding temperature dependence of DCP of the HoBL. }
	\label{Figure_11}
\end{figure}

The temperature-dependent valley polarization of the HoBL was measured under the same conditions as for the HBL (720 nm pulsed laser, $\sigma^+$ excitation) shown in Figure\ref{Figure_11}(a). The corresponding PL spectra and DCP are shown in Figure \ref{Figure_11}b). The analysis follows the same procedure as for the HBL; however, the DCP values are very low, indicating that these excitons exhibit negligible valley polarization in the HoBL

\newpage

\section{Supplementary Note 8: Time-resolved Photoluminescence of Phonon-assisted Intervalley Excitons in the HoBL}

\begin{figure}[h]
	\centering\includegraphics[width=15cm]{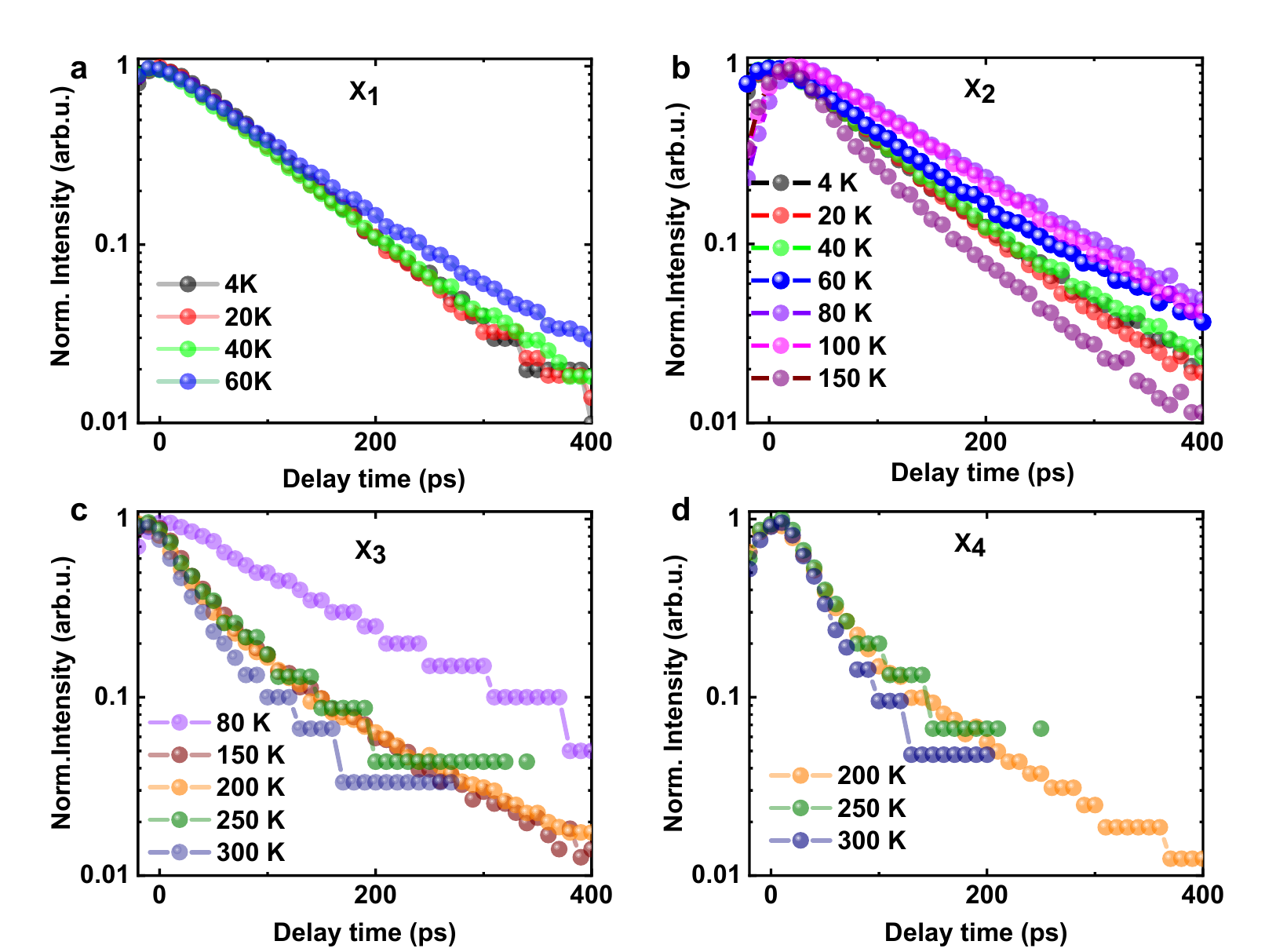} 
	\caption{\text{TRPL measurements of photon assisted intervalley excitons of HoBL.} \textbf{a-d} normalized PL intensity time trace of phonon assisted exciton X\textsubscript{1}, X\textsubscript{2}, X\textsubscript{3}, and X\textsubscript{4} in semi-logarithmic scale   }
	\label{Figure_12}
\end{figure}

The recombination dynamics of the phonon-assisted intervalley excitons in the HoBL were investigated using TRPL measurements under pulsed laser excitation at 720 nm. The TRPL decay curves were recorded as a function of temperature and plotted on a semi-logarithmic scale in Figure \ref{Figure_12}a-d) to analyze the exciton lifetimes. These measurements allow us to probe the temperature-dependent relaxation pathways of those four intervalley excitons.
\newpage

\section{Supplementary Note 9: Temperature-dependent Photoluminescence Characteristics of the HTL}

\begin{figure}[h]
	\centering\includegraphics[width=15cm]{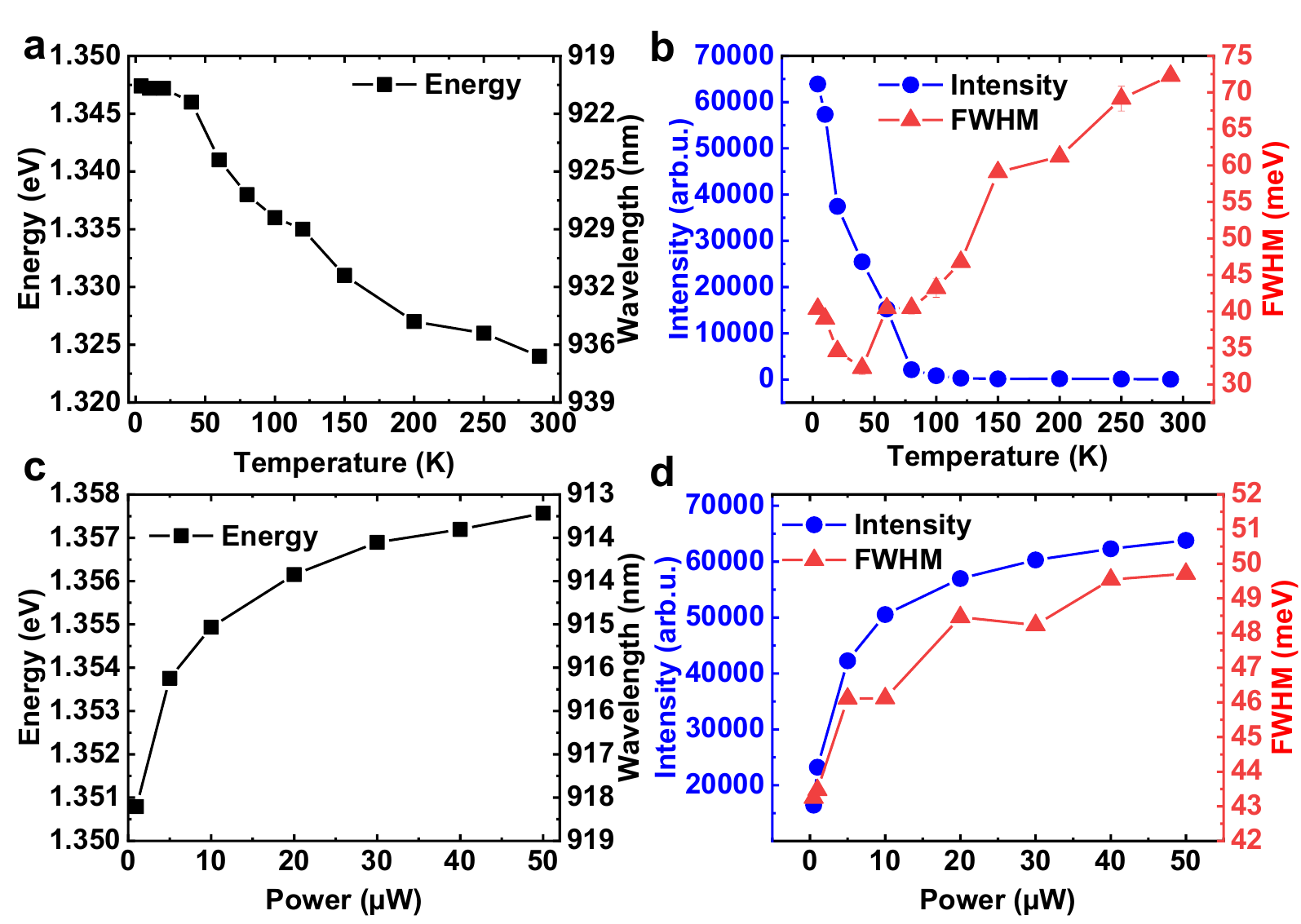} 
	\caption{\text{Temperature- and power-dependent PL characteristics of the HTL.} \textbf{a} Variation of HTL IX emission as a function of temperature. \textbf{b} Temperature-dependent variation of HTL IX intensity and FWHM. \textbf{c,d} Power-dependent evolution of energy, intensity, and FWHM of HTL IX.}
	\label{Figure_13}
\end{figure}

The temperature- and power-dependent PL response of the HTL exhibits trends similar to those observed in the HBL, as described above. Under pulsed excitation at 720 nm, the IXs emission in the HTL shows a systematic evolution of energy, intensity, and FWHM with temperature and excitation power (Figure \ref{Figure_13}a,b). Notably, the PL intensity of the HTL demonstrates a more pronounced temperature dependence, reflecting a stronger sensitivity of its excitonic emission to thermal effects. The power-dependent behavior follows the same general trends as in the HBL, indicating comparable exciton–exciton interaction dynamics in the HTL (Figure \ref{Figure_13}c, d)).

\section{Supplementary Note 10: Temperature-dependent Valley Polarization of the HTL}

\begin{figure}[h]
	\centering\includegraphics[width=15cm]{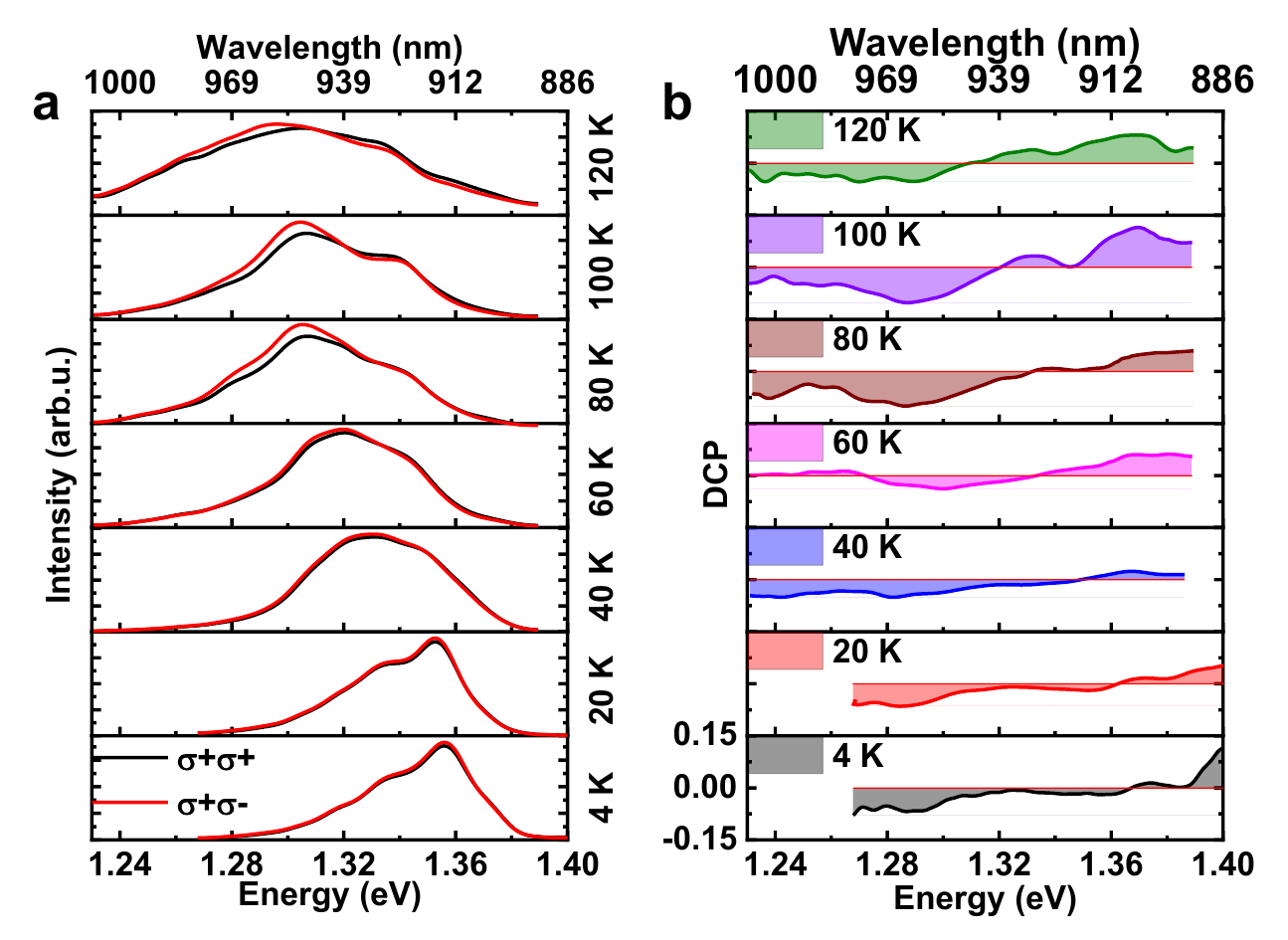} 
	\caption{\text{Temperature-tunable valley polarization of the HTL.} \textbf{a} Temperature-dependent $\sigma^{+}$ and $\sigma^{-}$ emission components acquired under $\sigma^{+}$ circular excitation, showing the evolution of valley selective optical response. \textbf{b} Extracted DCP as a function of temperature.}
	\label{Figure_14}
\end{figure}

The valley polarization of the HTL was measured under the same experimental conditions as for the HBL studies, using a pulsed 720 nm right circularly polarized ($\sigma^+$) excitation displays in Figure \ref{Figure_14}a). While the measurement procedure is identical to that described for the HBL, the temperature-dependent PL and DCP evolution of the HTL as observed in Figure \ref{Figure_14}b) reveals a distinctly different behavior. Unlike the HBL, which exhibits a switching between singlet and triplet states, the HTL maintains a broad singlet dominated emission that continuously redshifts with increasing temperature. This indicates that the HTL preserves a stable singlet IXs state over the measured temperature range, with a lower energy compared to the corresponding HBL states, reflecting modified spin–orbit coupling in the HTL configuration.

\section{Supplementary Note 11: Photoluminescence Evidence of Ultrafast IX Dynamics in the HTL}

\begin{figure}[h!]
	\centering\includegraphics[width=15cm]{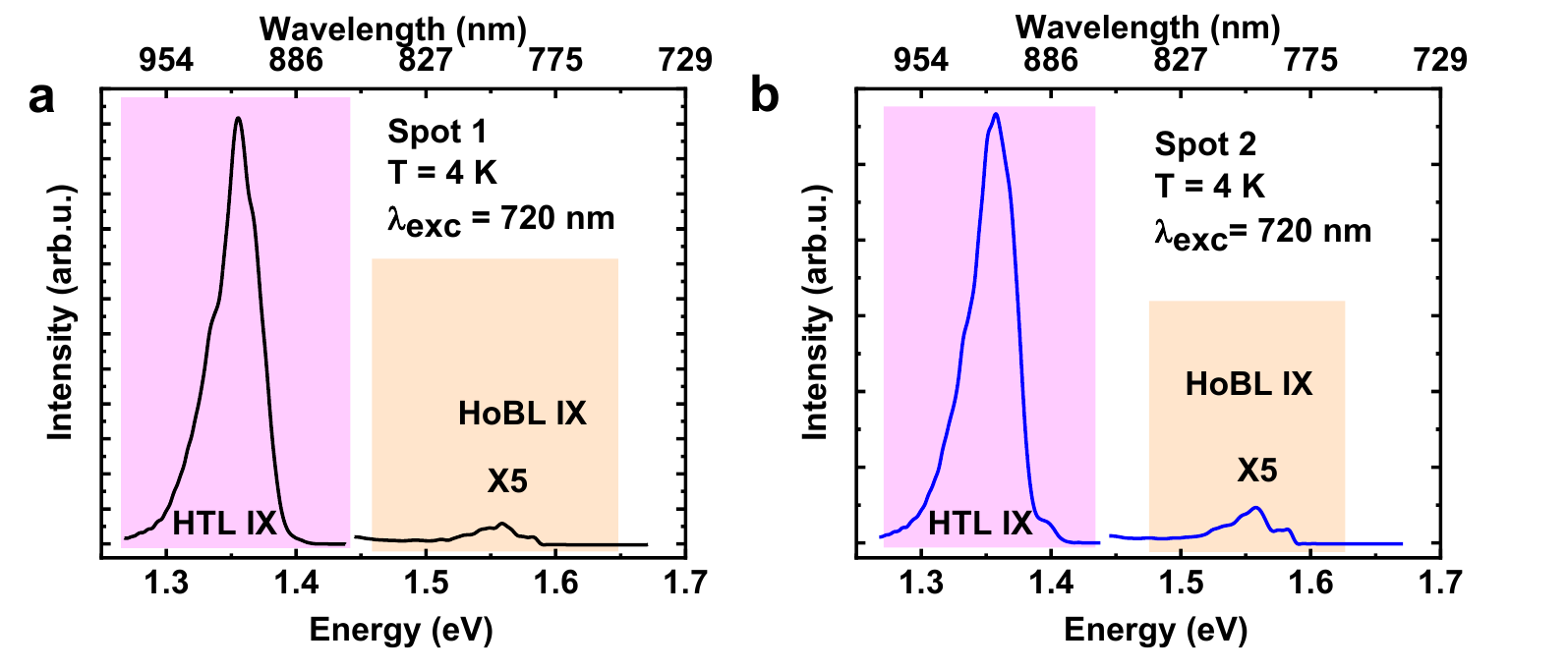} 
	\caption{\textbf{a-b} {Emission comparison from two spatial spots containing contributions from both HTL and HoBL regions at 4 K}. The HoBL intensity is magnified by a factor of five for visibility, yet remains significantly weaker compared to the HTL IX emission.}
    \label{Figure_15} 
\end{figure}

Spatially resolved PL from two regions containing HTL IXs coexisting with HoBL excitons, as shown in Figure \ref{Figure_15}. The PL intensity of the HTL IXs is significantly higher than that of the HoBL exciton emission, resulting in a strong quenching of the bilayer signal. This observation reflects an ultrafast IX transfer in the HTL, indicating efficient exciton funneling and energy redistribution from the HoBL to the HTL region.

\newpage

\section{Supplementary Note 12: Comparison of Photoluminescence Spectra at 4~K Across Four Spatially Distinct Measurements Spots}

\begin{figure}[h!]
	\centering\includegraphics[width=15cm]{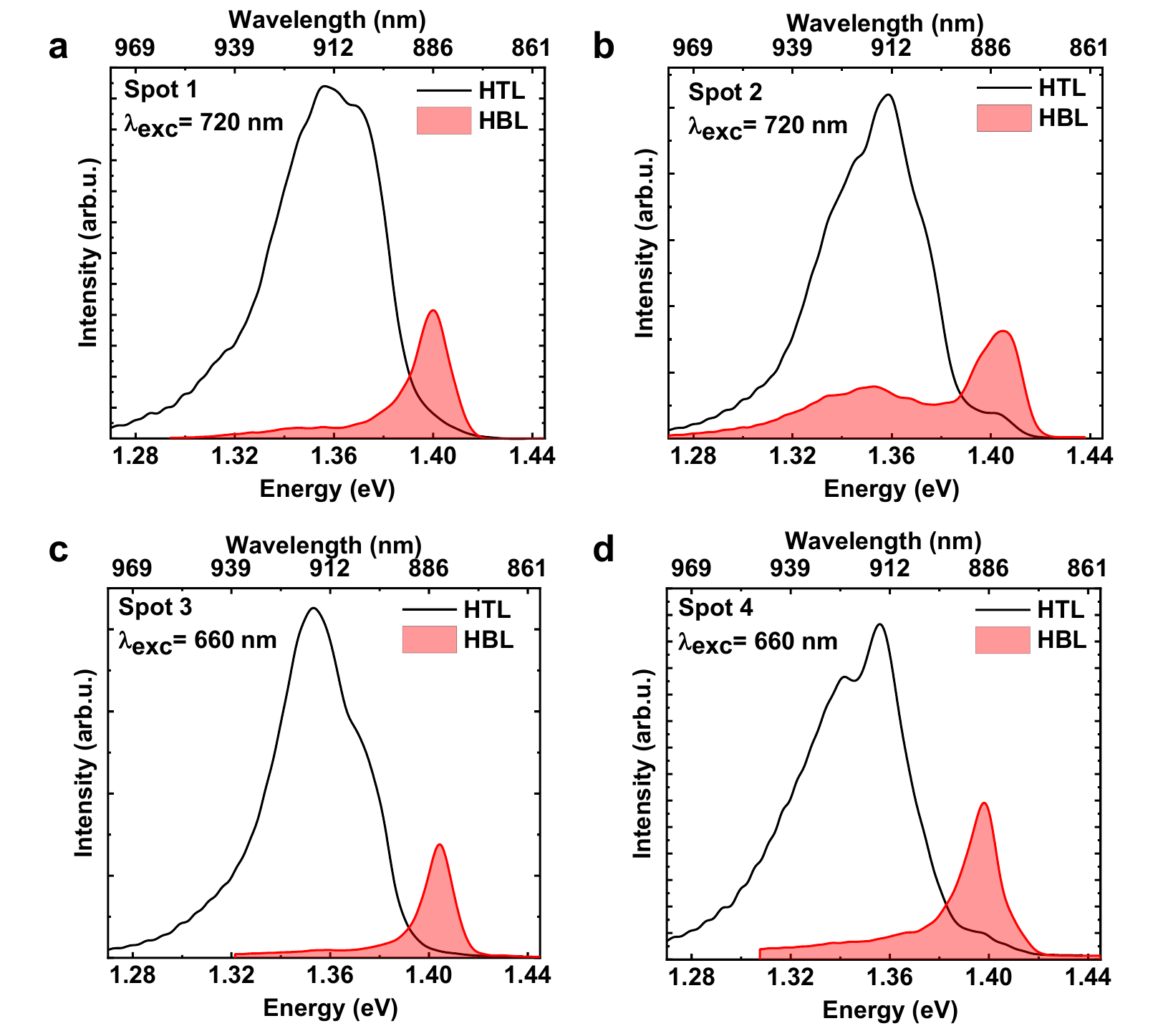} 
	\caption{\text{Comparison of PL spectra at 4~K from four spatially separated measurement spots.} \textbf{a-d} Each position reveals the characteristic emission features of the HTL and HBL, enabling evaluation of spectral uniformity and IX behavior across the sample.}
	\label{Figure_16}
\end{figure}

In addition to the spot discussed in the main text, four spatially separated positions of each HS (HBL and HTL) were investigated to assess the spectral uniformity and IX behavior of across the HS region. For each set, the first two spots were excited using a 720 nm laser (Figure \ref{Figure_16}a-b)), while the remaining two were excited with a 660 nm laser (Figure \ref{Figure_16}c-d)). Across all positions, the HTL consistently exhibits a higher PL intensity compared to HBL, highlighting enhanced IX emission in the HTL. 

\newpage

\section{Reference}
\bibliography{sorsamp}